\documentclass[%
 reprint,
 amsmath,amssymb,
 aps,
]{revtex4-1}
\usepackage{xcolor}
\usepackage{mathtools}
\usepackage{enumitem}
\usepackage{graphicx}
\usepackage{dcolumn}
\usepackage{bm}

\begin{document}

\title{Role of interactions in a closed quenched system}

\author{Bipasha Pal}
\author{Arvind Kumar Gupta}%

\affiliation{%
 Department of Mathematics, Indian Institute of Technology Ropar, Rupnagar, Punjab-140001, India}




\date{\today}
\begin{abstract}
We study the non-equilibrium steady states in a closed system consisting of interacting particles obeying exclusion principle with quenched hopping rate. Cluster mean field approach is utilized to theoretically analyze the system dynamics in terms of phase diagram, density profiles, current, etc with respect to interaction energy $E$. It turns out that on increasing the interaction energy beyond a critical value, $E_c$, shock region shows non-monotonic behavior and contracts until another critical value $E_{c_1}$ is attained; a further increase leads to its expansion. Moreover, the phase diagram of an interacting system with specific set of parameters has a good agreement with its non-interacting analogue. For interaction energy below $E_c$, a new shock phase displaying features different from non-interacting version is observed leading to two distinct shock phases. We have also performed Monte Carlo simulations extensively to validate our theoretical findings.

\end{abstract}

\maketitle

\vspace{2cm}
\section{\label{sec:level1}Introduction}

Driven diffusive systems, owing to their occurrences in large number of physical and biological processes, are of great significance. Some of the familiar experiences such as the flocking of birds or fish \cite{spector2005emergence}, ant trails \cite{schadschneider2010stochastic,chowdhury2005physics}, traffic flow \cite{schadschneider2010stochastic,chowdhury2005physics}, or biological transport \cite{schliwa2006molecular,macdonald1969concerning}, are just a few examples of such systems. These systems fall into non-equilibrium category which is far less understood than the equilibrium counterpart. However, the systems can settle down to a non-equilibrium steady state (NESS) which can be studied to gain deep insights into their properties. One of the most powerful tools in investigating multi-particle non-equilibrium system is a class of models called the Totally Asymmetric Simple Exclusion Process (TASEP) \cite{chou2011,sarkar2014nonequilibrium,kolomeisky1998phase,karzig2010signatures,kolomeisky2015motor}. It was originally proposed as a simple model for the motion of multiple ribosomes along mRNA during protein translation \cite{macdonald1968kinetics}. The model involves hopping of particles from one site to immediate next site on a one-dimensional lattice with a unit rate and obeying hardcore exclusion principle \cite{macdonald1968kinetics,derrida1998exactly,kolomeisky1998phase}. Over the years, due to its simplicity, different versions of TASEP have been extensively employed in studies of various aspects of biological motors, vehicular traffic,etc. providing important insights into these complex processes \cite{zia2011modeling,arita2015exclusive}.

Various versions of TASEP models have been thoroughly investigated. With open boundaries and unit hopping rates, the system settles into one of the three phases depending upon entry rate $(\alpha)$ and exit rate $(\beta)$. These  phases are referred to as high density (HD), low density (LD) and maximal current (MC) \cite{kolomeisky1998phase}. A variant of this system incorporated with weak correlations has a topologically similar phase diagram \cite{teimouri2015theoretical}. In a closed system with unit hopping rates, depending upon the number of particles, HD, LD and MC is obtained \cite{derrida1998exactly,chowdhury2005physics,nagel1992cellular}. With the particles interacting with energy $E$, the system exhibits a higher value for maximal current in case of weak interactions \cite{celis2015correlations}. The unit hopping rates, considered for simplicity, generally do not hold true for majority of systems \cite{malgaretti2012running,antal2000asymmetric,chowdhury2005physics,banerjee2020smooth}. For instance, a vehicle on a road can move with non uniform speed; it may slow down when it encounters a sharp turn or a speed breaker, etc., or its speed may be altered due to different speed limits for different parts of the road. Also, in simple TASEP, the particles are non interacting, i.e., hopping rate at a site is constant and remains unaffected by the occupancy of neighboring sites. However, in vehicular traffic it is observed that a vehicle slows down in presence of a vehicle immediately ahead of it, and speeds up if a vehicle behind it starts honking \cite{antal2000asymmetric}. Thus particles are influence by the presence of another particle. Similarly, experimental studies on kinesin motor proteins, which move along microtubules, indicate that these molecular motors interact with each other \cite{roos2008dynamic}. A model for kinesin having nearest neighbor particle interactions has been studied wherein the hopping rates modified due to interactions taking into consideration the fundamental thermodynamic consistency \cite{celis2015correlations,midha2018effect}. A recent study considers a system consisting of non-interacting particles on a closed lattice with quenched hopping rates i.e., hopping rate depends upon site, and shows the dependence of NESS on total number of particles and minimum of the quenched rates. The minimum acts as bottleneck and a shock in form of localized domain wall is observed in maximal current phase \cite{banerjee2020smooth}.

Motivated by the relevance of both quenched hopping rate and particle-particle interactions, we incorporate them in the simple TASEP to analyze a generalized version. Taking the recent studies into account, our objective is to answer the following questions. Does the simple mean field theory, which worked for the system of non-interacting particles \cite{banerjee2020smooth}, give accurate predictions? If no, then what advanced theory can be applied to obtain the dynamics of such a system? Do the qualitative properties of the system change with the inclusion of the interactions? How does the incorporation of interactions affect the shock phases? We proceed to answer these questions and a few more in the forthcoming sections using various mean-field analysis.
\section{\label{sec:level2}Theoretical Description}
\begin{figure}[h]
\centering
\includegraphics[width=.465\textwidth,totalheight=0.15\textheight,trim=4cm 17.7cm 4.55cm 5.24cm,clip=true]{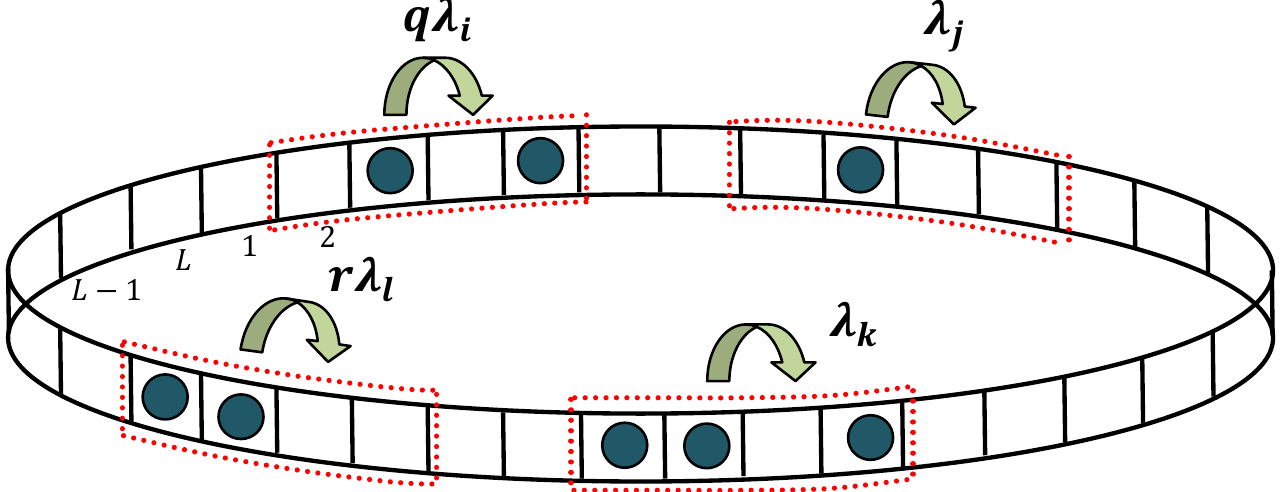}
  \caption{Schematic diagram of the closed TASEP with site dependent hopping rate that are modified according to occupancy of nearest neighbors.}
  \label{fig:model1}
\end{figure}

We consider a model comprised of $N$ interacting particles on a closed lattice with $L$ sites. Particles move unidirectionally on the lattice and obey the exclusion principle i.e., no two particles can occupy a single site. A particle hops to the next site only if the target site is empty. Unlike the unit hopping rate in simple TASEP, we consider quenched hopping rates characterized by $\lambda_i$ for $i^{th}$ site. This incorporation takes into account the inhomogeneity of the paths i.e., twists and turns in microtubules, and speed breakers, speed limits, etc., in roads and thus, makes the model more applicable. Additionally, the particles in the system interact with energy $E$ which is associated with the bonds connecting two nearest neighboring particles, where $E>0$ and $E<0$ corresponds to attraction and repulsion, respectively \cite{celis2015correlations,teimouri2015theoretical}. When a particle hops to its next site, it can break or form a bond depending on occupancy of neighboring sites that can be interpreted as opposite chemical reactions. For attractive interactions ($E>0$), a particle has a tendency to form a bond with the particle ahead of it, thereby increasing its rate of hopping by a factor $q(>1)$; whereas a particle resists the breaking off from bond with the particle behind it, leading to a decrease in its rate of hopping to next site by a factor $r(<1)$. Similar arguments hold for the repulsive interactions ($E<0$). The rates $q$ and $r$ are taken in a thermodynamic consistent manner which defines the forming and breaking of particle-particle bond as $q = e^{\theta E}$ and $r = e^{(\theta-1)E}$, respectively, where $\theta$ $(0<\theta\leq1)$ allows the measure of the distribution of energy \cite{teimouri2015theoretical}. Throughout our paper, we assume $\lambda_i=\lambda(i/L)$ having $0<\lambda(i/L)\leq1$ where $\lambda(.)$ is spatially piecewise smooth and has a single point global minimum. Depending upon the occupancy of neighboring sites, the hopping rate from site $i-1$ to $i$  is defined as follows (see Fig(\ref{fig:model1})):
\begin{itemize}
  \item when site $i-2$ and $i+1$ are empty (or occupied), the hopping rate is $\lambda_i$.
  \item when site $i-2$ is empty and site $i+1$ is occupied, the hopping rate is $r\lambda_i$ with $r\neq1$.
  \item when site $i-2$ is occupied and site $i+1$ is empty, the hopping rate is $q\lambda_i$ with $q\neq1$.
\end{itemize}
In the absence of interactions i.e., $E=0$, we have $q=r=1$ and thus we recover the TASEP  for closed ring with site dependent hopping rate \cite{banerjee2020smooth}. Further, if $\lambda_i$\rq{}s$=1$, we get the simple closed TASEP \cite{derrida1998exactly}.

On labeling the sites by $x=i/L$, thermodynamic limit $N\rightarrow\infty$ impels $x$ to become a quasi continuous variable confined between $0$ and $1$. We denote the steady-state probability of an $n-$cluster $(\tau_i,\tau_{i+1},\dots,\tau_{i+n-1})$ by $P(\tau_i,\tau_{i+1},\dots,\tau_{i+n-1})$ where $\tau_i$ denotes occupancy of $i^{th}$ site. Then, the steady state current is given by
\begin{eqnarray}
\label{eq:particlecurrent}
J&=&\lambda(x)P(0,1,0,0)+q\lambda(x)P(0,1,0,1)\nonumber\\
&&+r\lambda(x)P(1,1,0,0)+\lambda(x)P(1,1,0,1)
\end{eqnarray}
The four point correlators in Eq.(\ref{eq:particlecurrent}) makes it intractable in the present form which, in turn, prompts us to look for an approximation to the correlators. We approach this problem with mean field theory. The basic premise is to break the four point correlators into smaller correlators. We begin with the simple mean field (SMF) approximation wherein the idea is to ignore all possible correlations between the particles and probability of products is replaced by products of probabilities, i.e.,
\begin{equation}
\label{eq:smfapprox}
P(\tau_{i}\tau_{i+1}\tau_{i+2}\tau_{i+3})\approx P(\tau_{i})P(\tau_{i+1})P(\tau_{i+2})P(\tau_{i+3})
\end{equation}
Since $\langle\tau_i\rangle=\rho(i/L)$, where $\rho(.)$ denotes particle density, we have $P(1)=\rho$ and $P(0)=1-\rho$. Under the approximation given by Eq.(\ref{eq:smfapprox}), Eq.(\ref{eq:particlecurrent}) becomes
\begin{equation}
\label{eq:Jrhoequation}
J=\lambda(x)\rho(1-\rho)(1+\rho^2(2-q-r)-\rho(2-q+r))
\end{equation}
 In contrast to the case of simple TASEP with interacting particles, here $\rho$ will not be a constant throughout the lattice \cite{midha2018effect}. Solving Eq.(\ref{eq:Jrhoequation}) leads us to the following expressions for density profile:
\begin{equation}
\label{eq:ldrho}
\rho_{\pm}(x)=\begin{cases}
\frac{1}{2} \Bigg[1 \pm \sqrt{ 1+ \frac{2 (1- \sqrt{\frac{4J( q +  r-2)}{ \lambda(x)} +1})}{ (
      q + r -2 )}}\Bigg], & q,r\neq 1\\
      \\
      \frac{1}{2}\Bigg[1\pm\sqrt{ 1-\frac{4J}{\lambda(x)}}\Bigg],& q=r=1
      \end{cases}
\end{equation}
for all $x$. Clearly, $\rho_-(x)$ ($\rho_+(x)$) is bounded above (below) by 0.5, and depends upon $J$. This $J$ can be calculated by using the particle number conservation (PNC):
\begin{equation}
\label{eq:pnc}
\int_0^1\rho_a(x)=n, ~~~~~~a=+,-
\end{equation}
where $n=\frac{N}{L}$. For feasible values of $\rho(x)$,

 \begin{equation*}
 J\leq\begin{cases}
      \dfrac{(2+q+r)\lambda(x)}{16}, &q,r\neq 1\\
     \dfrac{\lambda(x)}{4},& q=r=1
     \end{cases}
      \end{equation*}
      for all $x$. Thus, the maximum possible value of particle current is
       \begin{equation}
       \label{eq:Jmax}
      J_{max}=\begin{cases}\dfrac{(2+q+r)\lambda_{min}}{16}, &q,r\neq 1\\
      \dfrac{\lambda_{min}}{4},& q=r=1 \end{cases}
      \end{equation}
         where $\lambda_{min}$ is the global minimum of the $\lambda(x)$. At  $J=J_{max}$, $\rho_{-}(x_0)=\rho_{+}(x_0)$ where $x_0$ is the point of global minimum. For the limiting case, as $E \rightarrow 0$, the expressions for density profile and maximal current agrees well with that obtained in Ref.\cite{banerjee2020smooth}. Further if $\lambda(x)=1$, the results exactly match with that of the simple TASEP with non-interacting particles.
\begin{figure}[h]
         \centering
         \includegraphics[height=1.6in]{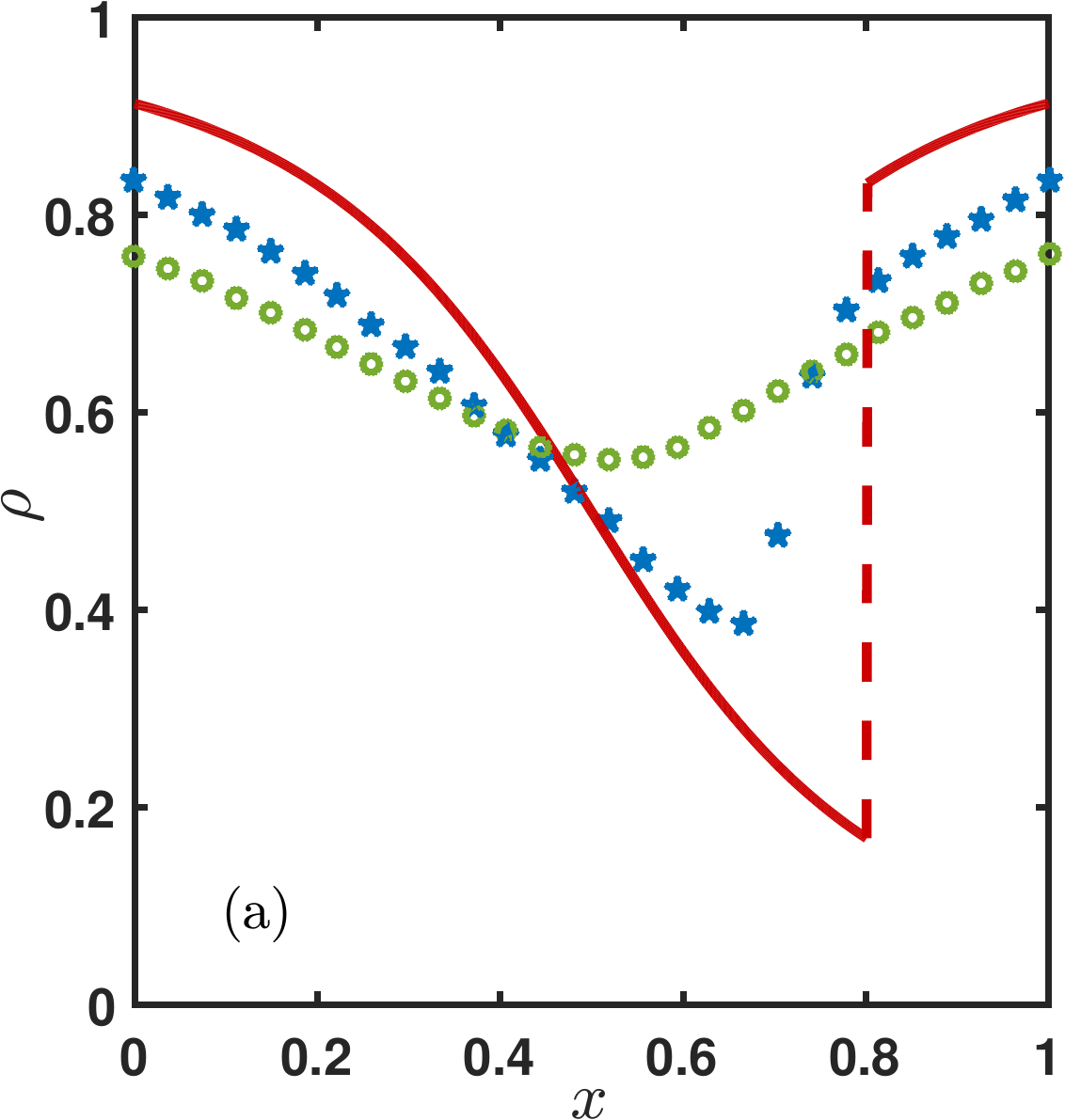}
         \includegraphics[height=1.59in]{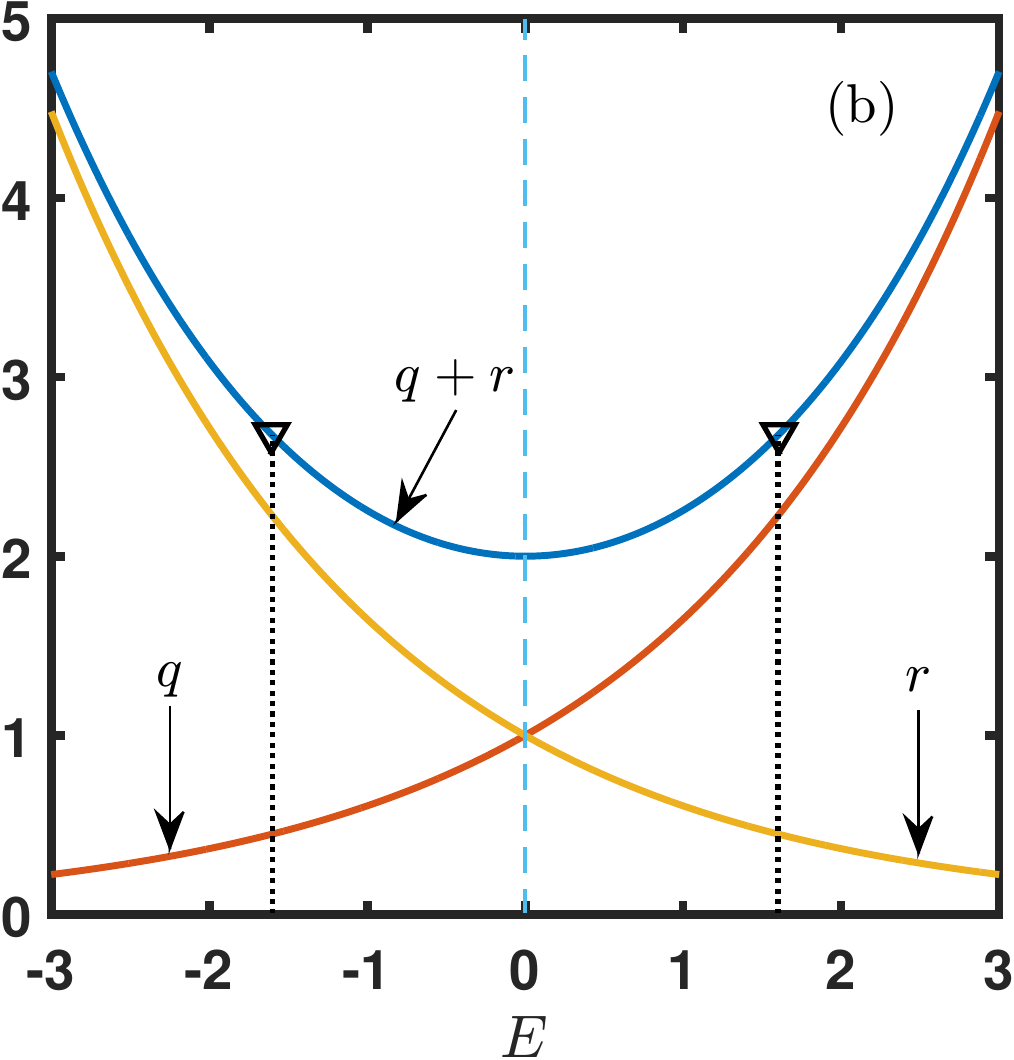}
        \caption{(a) Density profile: SMF results (denoted by red dashed line) for $E=1.6$ and $E=-1.6$ and $n=0.65$, MCS results (denoted by symbols) for $E=1.6$ and $n=0.65$ (hexagons) and $E=-1.6$ and $n=0.65$ (circles) where $\lambda(x)=(x-0.5)^2-0.5$  (b) Graph showing $q+r$ is even function. Blue line, yellow line and red line denote $q+r$, $q$ and $r$ respectively. Black inverted triangles denote value of $q+r$ at $E=-1.6$ and $E=1.6$ which turn out to have same values.}
        \label{fig:qre}
        \end{figure}

\begin{figure}[h]
    \centering
       \includegraphics[height=1.5in]{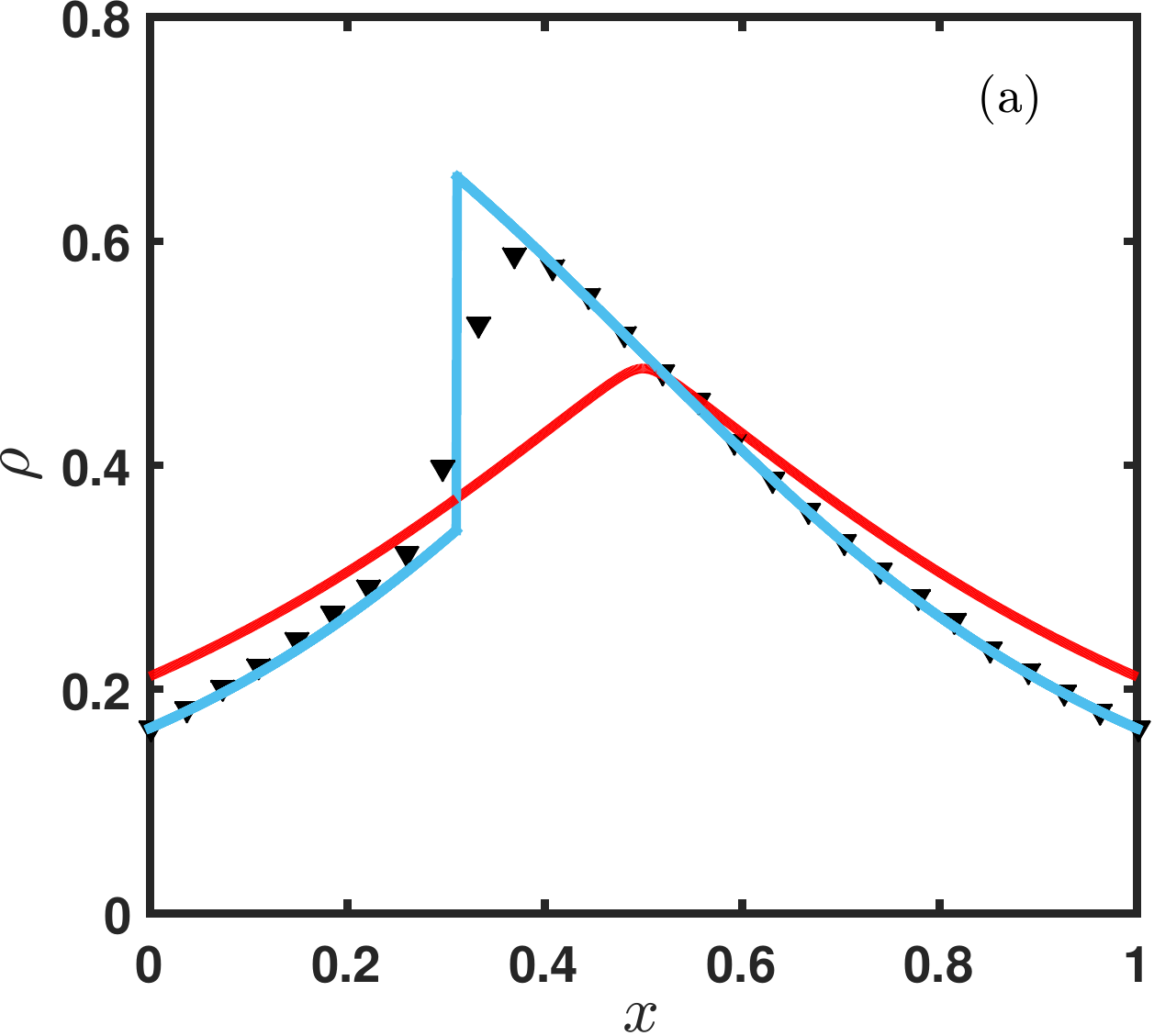}
       \includegraphics[height=1.5in]{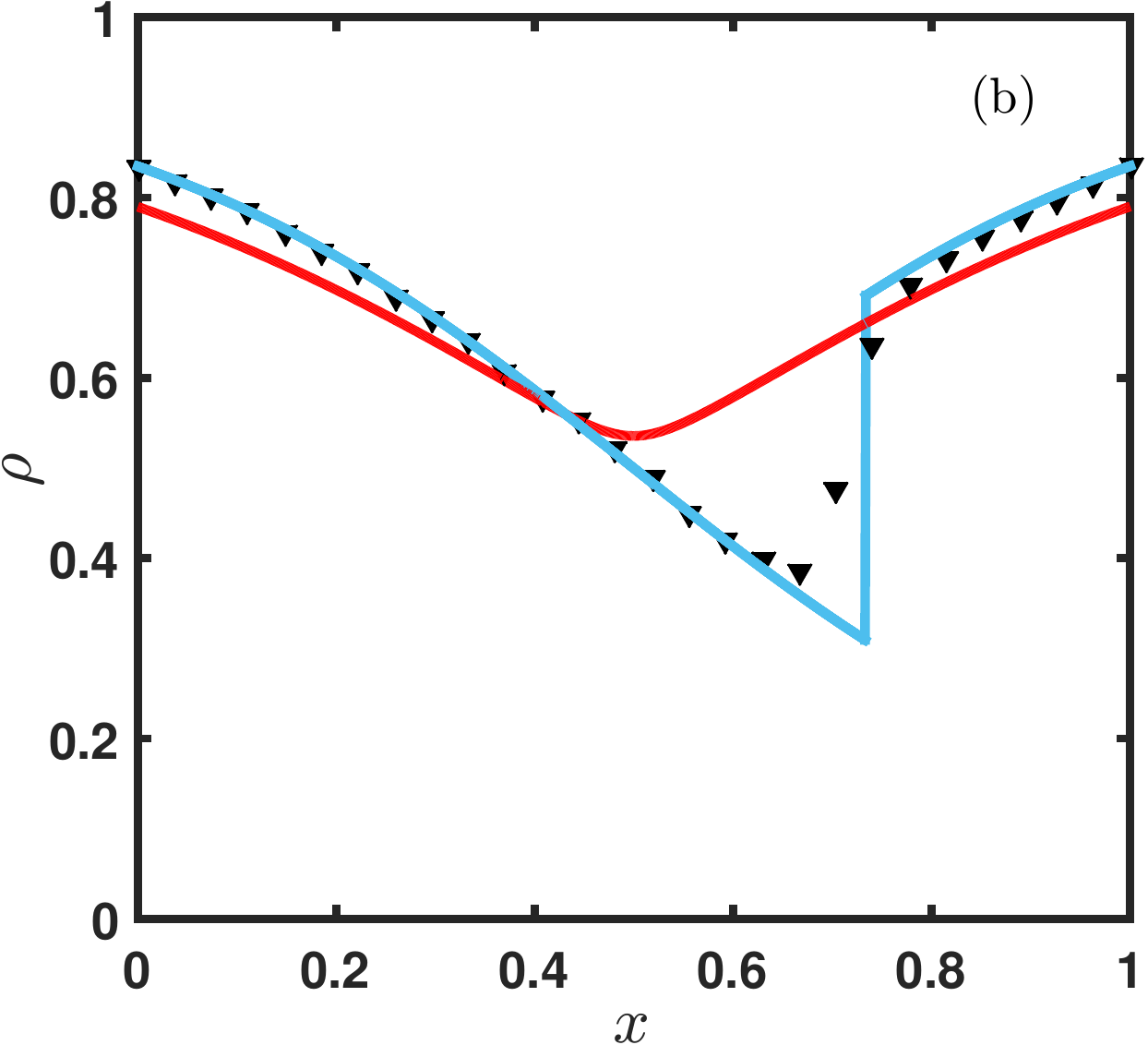}\\
      \includegraphics[height=1.5in]{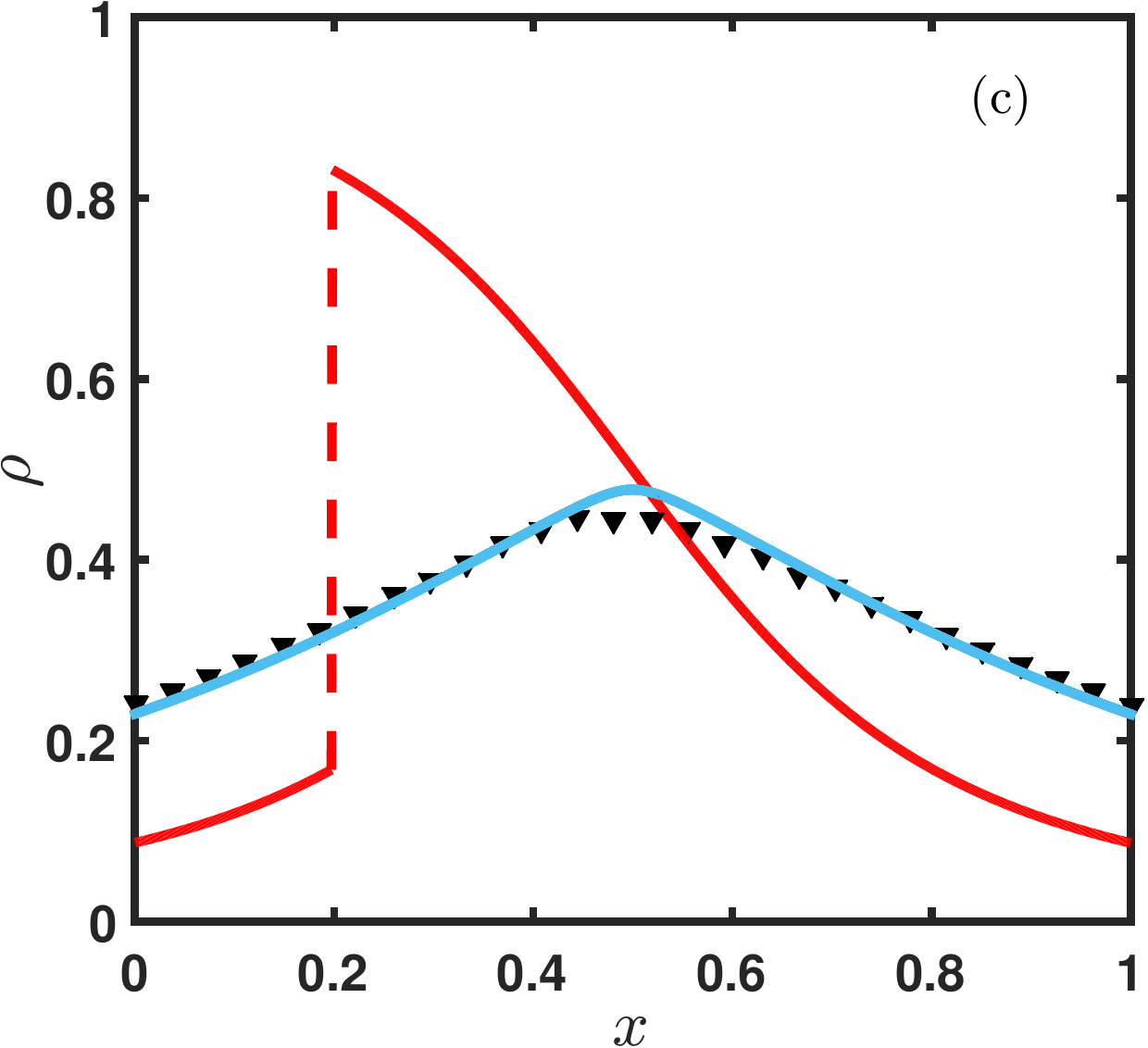}
      \includegraphics[height=1.5in]{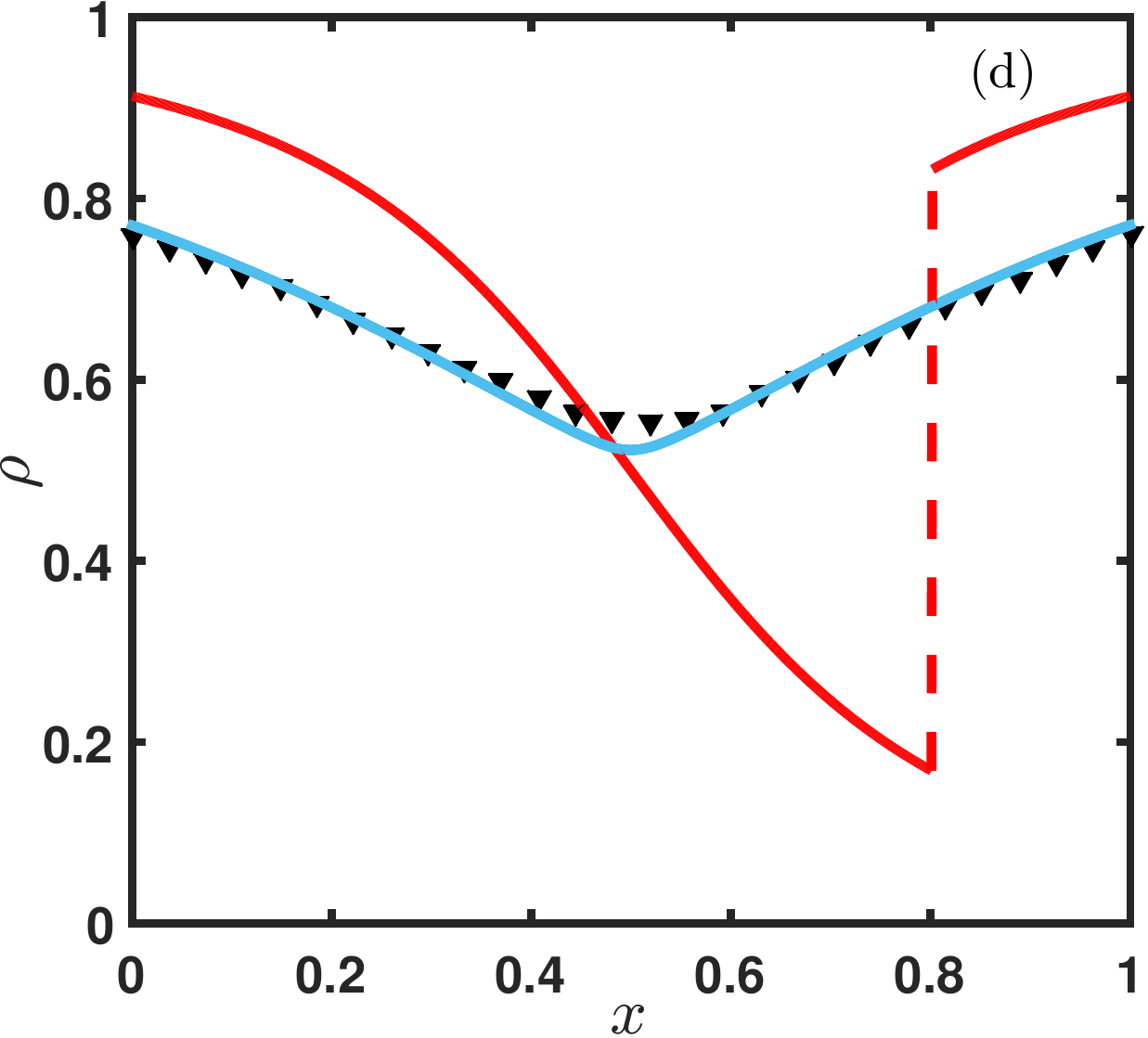}
    \caption{Density profiles  for $\lambda(x)=(x-0.5)^2-0.5$: (a) $E=1.6$, $n=0.34$, (b) $E=1.6$, $n=0.66$, (c) $E=-1.6$, $n=0.35$, (d) $E=-1.6$, $n=0.65$,  Dashed line (red), solid line (blue) and symbols (black) shows SMF, CMF and MCS results, respectively. }
    \label{fig:comparisonmcssmf}
\end{figure}

The density obtained from SMF produces exactly the same profiles qualitatively as well as quantitatively for equal strength of attractive as well as repulsive interactions (see Fig.(\ref{fig:qre}(a)). It is due to the fact that $q+r$ is an even function of $E$ (see Fig.(\ref{fig:qre}(b))). This finding differs largely when compared to Monte Carlo simulations (MCS) which shows that effect of attractive interaction differs from that of repulsive interactions for any interactive strength (see Fig.(\ref{fig:comparisonmcssmf}(a)).

 For different values of interactions, the maximal current predicted from SMF drastically varies from MCS (see Fig.4(a)). According to SMF, maximal current increases with increase in strength of interactions which is physically impossible. Clearly, SMF fails to predict the density profile and current accurately.  The correlation profile shows that the system has correlations which accounts for the failure of SMF \cite{article}(see Fig.4(b)). Therefore, we need to consider a modified version of SMF that incorporates some of the correlations.
\begin{figure}[h]
    \centering
    \includegraphics[height=1.4in]{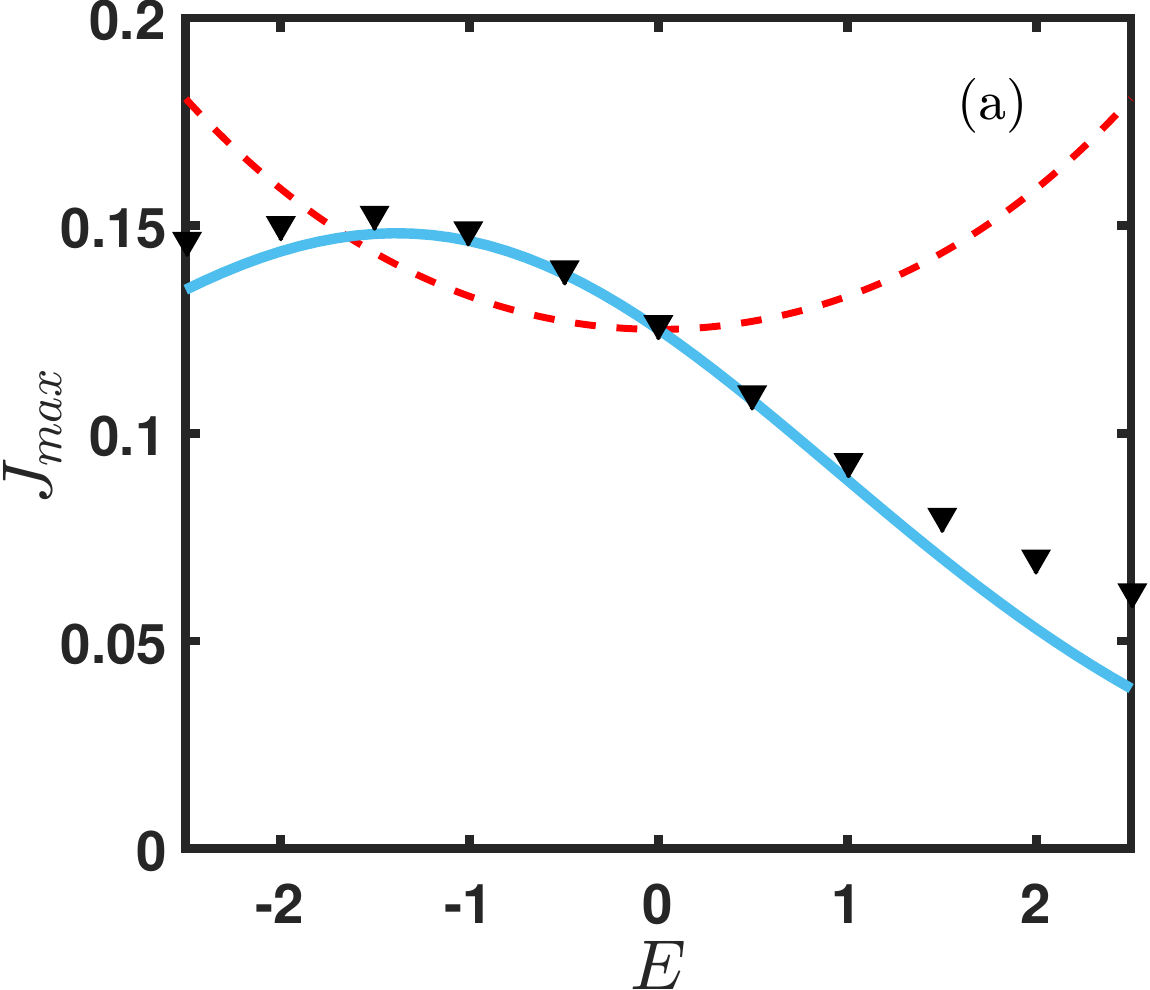}
       \includegraphics[height=1.4in]{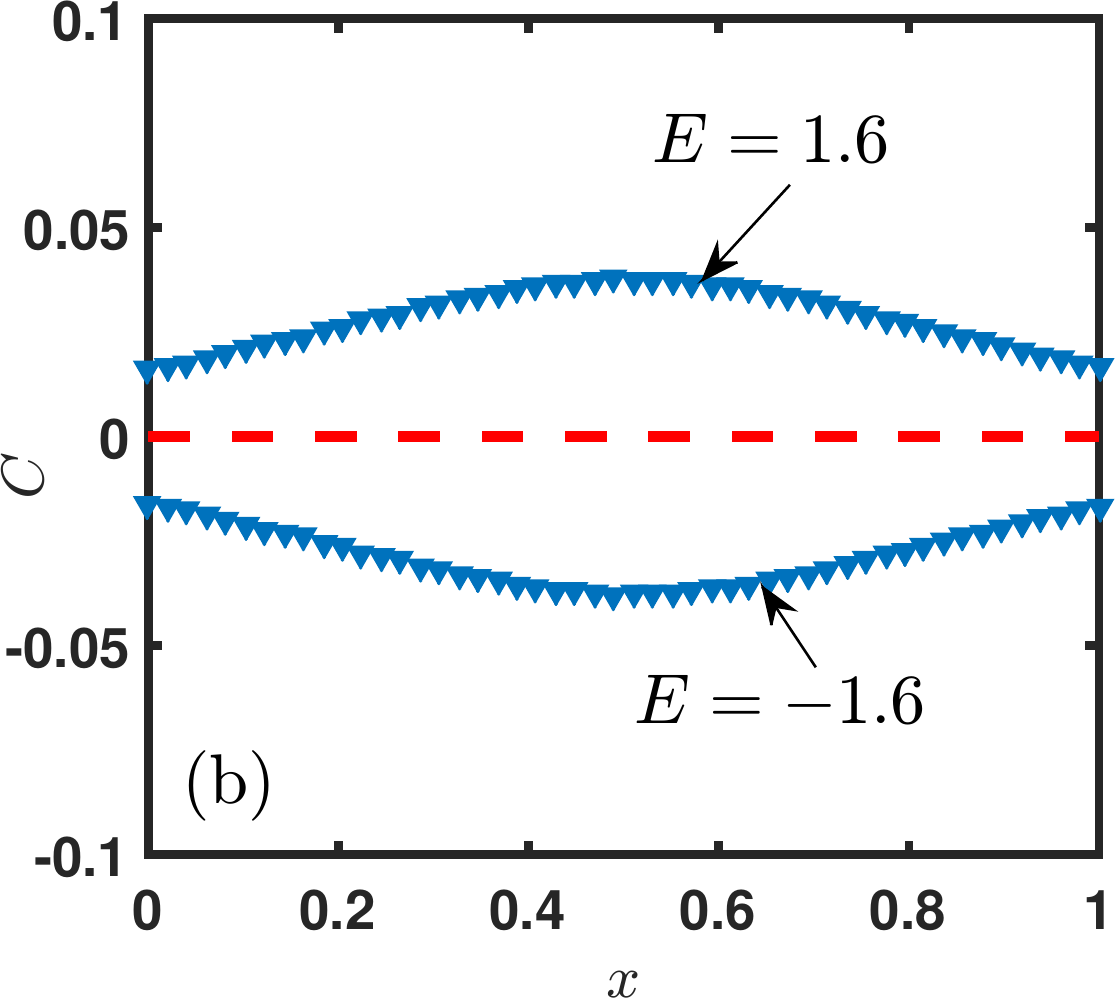}

    \caption{(a) Maximal current versus $E$ for  $\lambda(x)=(x-0.5)^2+0.5$. Dashed line (red), solid line (blue) and symbols (black) shows SMF, CMF and MCS results, respectively, (b) Correlation profile for $\lambda(x)=(x-0.5)^2+0.5$, MCS results are denoted by symbols and SMF result is denoted by dashed line.}

\end{figure}\\
To overcome the incapability of SMF for not handling interactions in closed lattice, we employ cluster mean field theory (CMF) which considers some correlations between nearest neighbors. Specifically, we use $2$-site CMF to factorize the probability of $n$-site cluster to product of probability of $2$-site cluster as follows:
\begin{equation}
P(\tau_1,\tau_2,...\tau_n)\propto P(\tau_1,\tau_2)P(\tau_2,\tau_3)...P(\tau_{n-1},\tau_n)
\end{equation}
which, after normalization,  gives
\begin{equation}
\label{eq:cmfapprox}
P(\tau_1,\tau_2,...\tau_n)=\frac{ P(\tau_1,\tau_2)P(\tau_2,\tau_3)...P(\tau_{n-1},\tau_n)}{P(\tau_2)P(\tau_3)...P(\tau_{n-1})}
\end{equation}
A probability of 2-cluster with two empty sites is labeled as $P(0, 0)$, with two occupied sites as $P(1, 1)$, and two half-occupied clusters are labeled as $P(1, 0)$ and $P(0, 1)$. Each 2-cluster can be found in any of four possible states. By particle-hole symmetry, $P(0,1)=P(1,0)$. Under CMF approximation given by Eq.(\ref{eq:cmfapprox}), the current-density relation is obtained as follows:
\begin{eqnarray}
\label{eq:cmfdensitycurrent}
J&=&\frac{\lambda(x)P(1,0)}{P(1)P(0)}\Big[P(1,0)\big(qP(1,0)+P(0,0)+P(1,1)\big)\nonumber\\
&&+rP(1,1)P(0,0)\Big]
\end{eqnarray}
By Kolmogorov conditions, we obtain
\begin{eqnarray}
\label{eq:rho}
P(1,0)+P(1,1)&=&\rho\\
\label{eq:1minusrho}
P(1,0)+P(0,0)&=&1-\rho
\end{eqnarray}
Further, steady state master equation for P(1,0) yields
\begin{equation}
\label{eq:cmfsteady01}
rP(1,1,0,0)-qP(0,1,0,1) = 0.
\end{equation}
Using Eq.(10-12), we obtain

\begin{eqnarray}
\label{eq:p11}
P(1,1)&=&\begin{cases}  \rho+\frac{r - \sqrt{r( r + 4 \rho(q-r)(1-\rho) )}}{2 (q - r)},\\ &q,r\neq 1\\
\rho^2,&q=r=1\end{cases}\\
\label{eq:p10}
P(1,0)&=&\begin{cases} \frac{-r+\sqrt{r( r + 4 \rho(q-r)(1-\rho) )}}{2 (q - r)},\\ &q,r\neq 1\\
\rho(1-\rho),&q=r=1\end{cases}\\
\label{eq:p00}
P(0,0)&=& \begin{cases} 1-\rho+\frac{r - \sqrt{ r( r + 4 \rho(q-r)(1-\rho) )}}{2 (q - r)},\\ &q,r\neq 1\\(1-\rho)^2,&q=r=1\end{cases}
\end{eqnarray}

The above equations suggest that $2$-site cluster probability doesn't depend explicitly on quenched hopping rates.
On solving Eq.(\ref{eq:cmfdensitycurrent}), (\ref{eq:p11}), (\ref{eq:p10}) and (\ref{eq:p00}) we obtain the following expression for steady state current:
\begin{eqnarray}
\label{eq:Jcmf}
J&=&\lambda(x)\Bigg(\frac{\sqrt{r^2+4r\rho(q-r)(1-\rho)}-r}{4(q-r)^3\rho(\rho-1)}\Bigg)\times\nonumber\\&&\Big(4r\rho(q-1)(q-r)(\rho-1)+(\sqrt{r^2+4r\rho(q-r)(1-\rho)}\nonumber\\&&-r)(2rq-q-r)\Big),
\end{eqnarray}
and thus the density profile can be obtained by
\begin{eqnarray}
\label{eq:rhocmfecrit}
\rho_{\pm}(x)&=&\frac{1}{2}\Bigg[1\pm \Bigg(1+ \frac{2}{4 r^3 \lambda(x)^2(q-1)^2} \bigg( J^2(r-q)^3\nonumber\\&&  +\> 2J\lambda(x)r(q-r)(q-3qr+2r)\nonumber\\&& -\>\lambda(x)^2r^2(r-q+2qr)\nonumber \\&&+\> \Big(J(r-q)^2 -\lambda(x)r(q+r-2qr) \big(J^2(r-q)^2 \nonumber\\ && -\>Jr\lambda(x)^2(2q-6r+8qr+r\lambda(x)\big)^{\frac{1}{2}}\Big)\bigg)\Bigg)^{\frac{1}{2}}\Bigg].
\end{eqnarray}
for $q,r\neq 1$. For $q=r=1$, we obtain
\begin{equation}
\label{eq:rhocmf}
\rho_{\pm}(x)= \frac{1}{2}\Bigg[1\pm\sqrt{ 1-\dfrac{4J}{\lambda(x)}}~\Bigg]
 \end{equation}
 Here $\rho_-(x)\leq 0.5$ and $\rho_+(x)\geq 0.5$ for all $x$.
The above expressions reduces to the results in Ref.\cite{midha2018effect} for $\lambda(x)=1$. We further explore the current for the extreme cases. For $E\rightarrow\infty$,  $P(1,1)\rightarrow\rho$ which leads to $J\rightarrow 0$. This is expected since particles form clusters due to large attractive energy which blocks their movement, whereas, for $E\rightarrow-\infty$, $P(1,1)\rightarrow 0$ which leads to $J\rightarrow\lambda\dfrac{\rho(1-2\rho)}{1-\rho}$ (see Appendix \ref{appendix:mapping}). When $\lambda(x)=1$, the expression matches with that of non interacting dimers \cite{midha2018theoretical}. Defining  $X=\rho(1-\rho)$ and substituting in Eq.(\ref{eq:Jcmf}) gives
\begin{eqnarray}
J&=&\lambda(x)\Bigg(\frac{r-\sqrt{r^2+4r(q-r)X}}{4(q-r)^3X}\Bigg)\Big(4r(1-q)(q-r)X\nonumber\\&&+\>(\sqrt{r^2+4r(q-r)X}-r)(2rq-q-r)\Big).
\end{eqnarray}
In order to obtain extrema of current-density relation, we determine $\rho$ such that
\begin{equation}
\label{eq:criticalpoints}
\frac{dJ}{d\rho}=\frac{dJ}{dX}(1-2\rho)=0.
\end{equation}
Clearly, $\rho=0.5$ is a critical point for all values of $E$. Other critical points exist if
\begin{equation}
\label{eq:criticalpoints2}
\Big(r+\sqrt{r^2+4r(q-r)X}\Big)^2=\frac{2r(q+r-2qr)}{1-q}.
\end{equation}
Substituting $X=0.25$ corresponding to $\rho=0.5$ in Eq.(\ref{eq:criticalpoints2}) yields
\begin{equation}
\label{eq:criticalenergy}
e^{\theta E}=\dfrac{1-e^\frac{E}{2}}{3+e^\frac{E}{2}}
\end{equation}
The above equation can be solved to obtain the critical interaction energy $E_c(\theta)$ for $0<\theta\leq1$. Note that $E_c(\theta)$ remains negative for all $0<\theta\leq  1$ which means that critical interaction energy is always repulsive in nature. For $E\leq E_c(\theta)$, Eq.(\ref{eq:criticalpoints2}) has two real roots. Therefore Eq.(\ref{eq:criticalpoints}) has only one critical point when $E>E_c(\theta)$, and three critical points when $E<E_c(\theta)$.  For $\theta=0.5$, Eq.(\ref{eq:criticalenergy}) gives $E_c=2\ln(\sqrt{5}-2)\approx-2.885k_BT$ which coincides with value obtained in \cite{midha2018effect,dierl2011time}. We further discuss these cases separately for $\theta=0.5$ since it splits the interaction energy symmetrically.
\subsection{$E>E_c$}
For interaction energy greater than $E_c$, $0.5$ is the only extreme point of current-density relation given by Eq.(\ref{eq:Jcmf}). Moreover, $J$ attains maximum value at $\rho=0.5$ which is given by
\begin{equation}
       \label{eq:cmfJmax}
      J_{max}=\begin{cases} \frac{\lambda(x_0)(\sqrt{qr}-r)\big(qr(q + r- 2) + \sqrt{qr}(q + r-2qr)\big)}{(q-r)^3},\\ ~~~~~~~~~~~~~~~~~~~~~~~~~~~~~~~~~~~~~~~~~~~~~~~~~~~~q,r\neq 1\\
      \dfrac{\lambda(x_0)}{4},~~~~~~~~~~~~~~~~~~~~~~~~~~~~~~~~~~~~~~~~~~ q=r=1 \end{cases}
      \end{equation}
      where $\lambda(x_0)=\lambda_{min}$.
Fig.(4(a)) shows that the current which is obtained from above expression agrees well with MCS results and overcomes the drawback of SMF approach. Additionally, the density profiles computed using CMF and MCS are in well agreement. (see Fig.\ref{fig:comparisonmcssmf})).

\begin{figure}[h]
\centering
\includegraphics[height=2.36in]{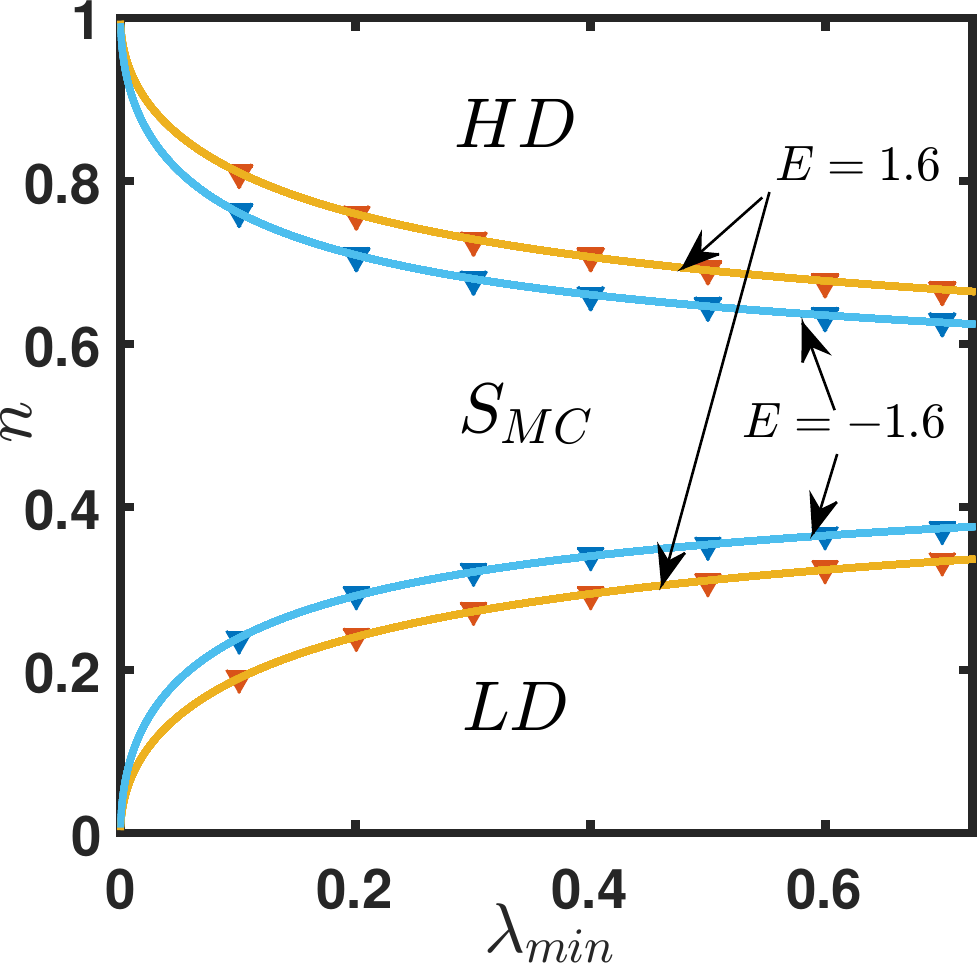}
\caption{ Phase diagram in $n-\lambda_{min}$ plane for $\lambda(x)=(x-0.5)^2+\lambda_{min}$, $E=1.6$ and $E=-1.6$. Solid lines and symbols denote CMF and MCS results, respectively.}
\label{fig:E_vs_lambdamin_phase_diagrams}
\end{figure}

We now investigate the effect of interactions on the phase digram in ($n-\lambda_{min}$) plane. It is observed that phase diagram obtained for the proposed model is qualitatively equivalent to that of a non-interacting system \cite{banerjee2020smooth} with three distinct phases: $LD$, $HD$ and $MC$ (see Fig.(\ref{fig:E_vs_lambdamin_phase_diagrams})). It has been found that MC phase has a jump discontinuity in density profile and hereafter we call it shock phase (denoted by $S_{MC}$).

\begin{figure}[h]
\centering
\includegraphics[height=2.6in]{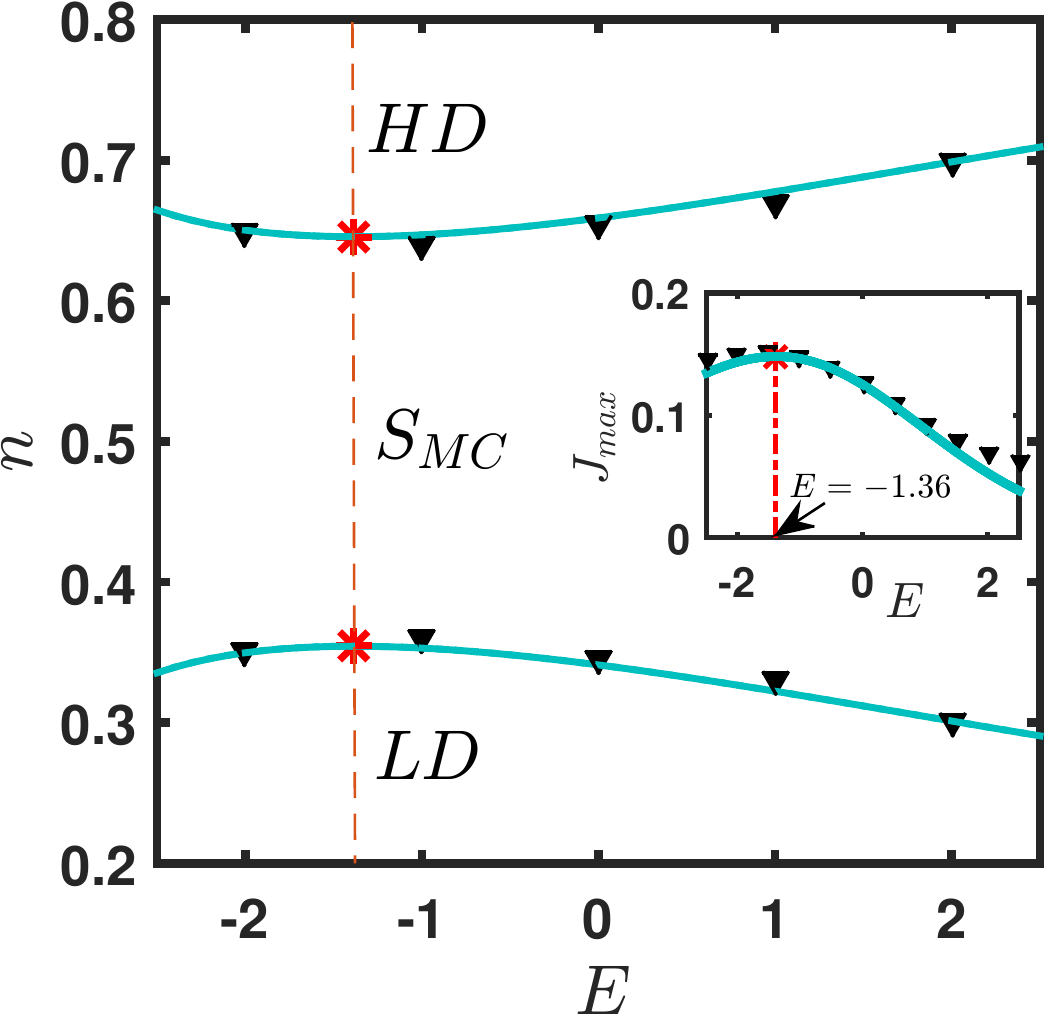}
\caption{ Phase diagram in $E-n$ plane  and (inset) variation of maximal current with respect to $E$ for $\lambda(x)=(x-0.5)^2+0.5$. Asterisk shows that shock region decreases until $E=-1.36$; thereafter it starts increasing. Solid lines and symbols denote CMF and MCS results, respectively.}
\label{fig:E_vs_n}
\end{figure}

\begin{figure}[h]
\centering
\includegraphics[height=1.4in]{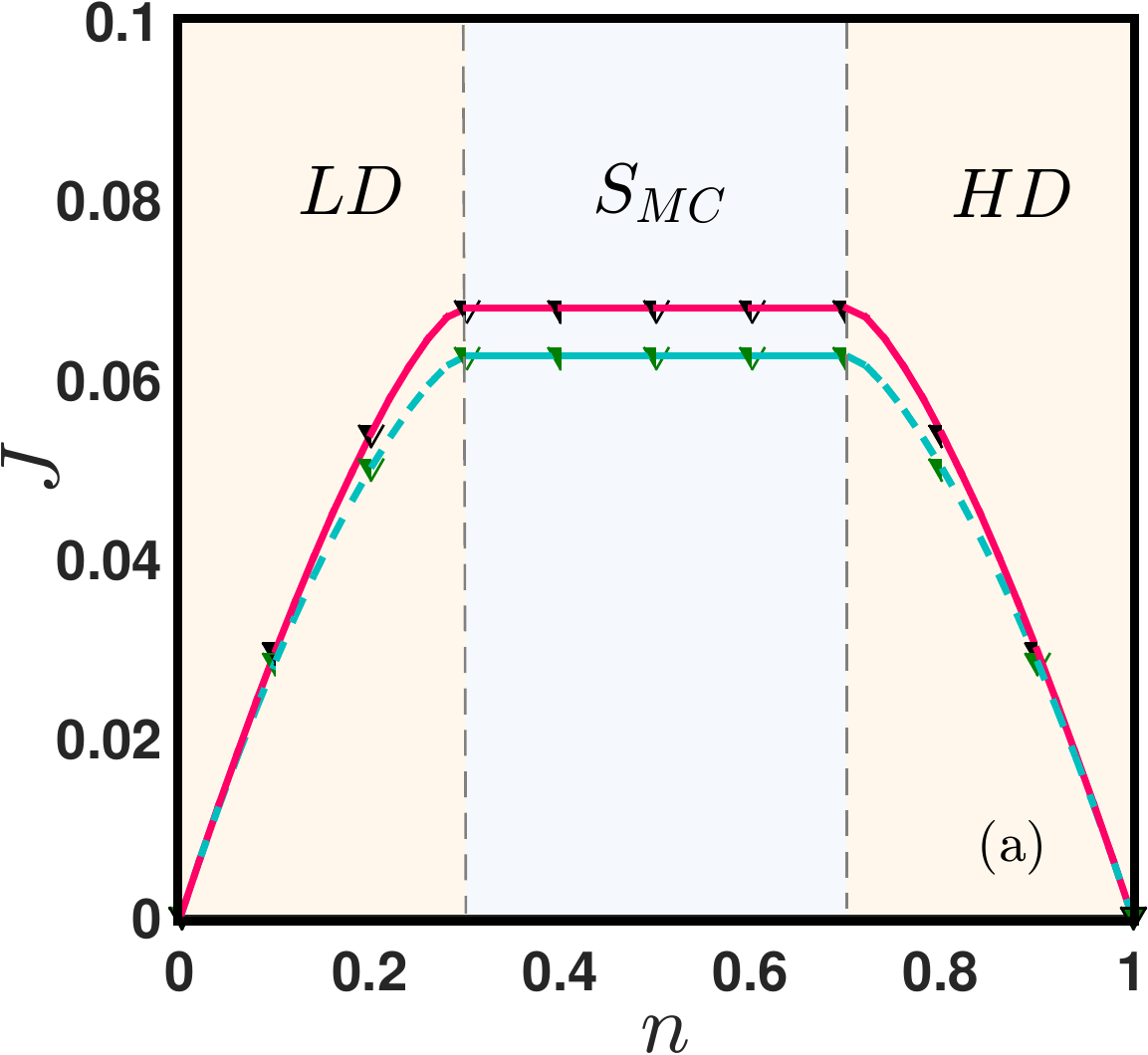}
\includegraphics[height=1.4in]{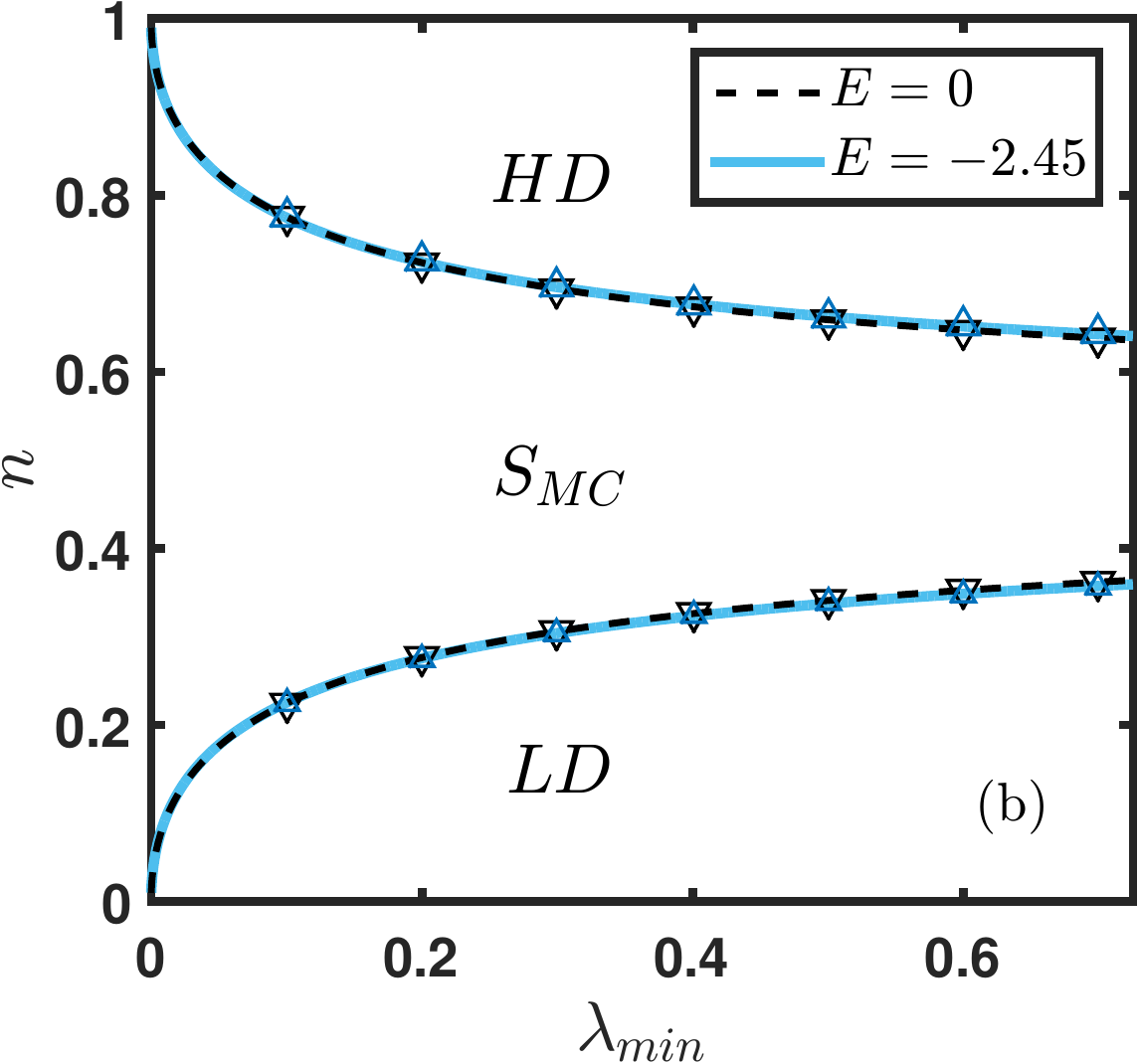}
\caption{(a) Current and phase line for $\lambda(x)=(x-0.5)^2+0.25$. Current for $E=0$ and $E=-2.45$ is denoted by dashed line and denotes solid line, respectively. (b) Phase diagram in $n-\lambda_{min}$ plane of $E=-2.45$ and $E=0$ for $\lambda(x)=(x-0.5)^2+\lambda_{min}$. Solid lines denote CMF results and symbols denote MCS results.}
\label{fig:phaseline}
\end{figure}

To comprehend the effect of varying the interaction energy, we first examine the behavior of maximal current. As $E$ increases, $J_{max}$ increases until it attains its highest value at a critical energy ($E_{c_1}$) followed by a continuous decrease (see inset, Fig.(\ref{fig:E_vs_n})). For $\lambda(x)=(x-0.5)^2+0.5$, $E_{c_1}\approx -1.36$ which is theoretically computed from Eq.(\ref{eq:cmfJmax}) and agrees well with that obtained by MCS. The boundary between phases $LD$ and $S_{MC}$, for a fixed $\lambda_{min} $, is given by
\begin{equation}
\label{eq:phaseboundary}
n_0=\int_0^1\rho_-\big(x,J_{max}\big)dx,
\end{equation}
Similar arguments hold for the phase boundary between $S_{MC}$ and $HD$. Thus, as we increase $E$, the shock phase shrinks until $J_{max}$ reaches its maxima, thereafter it starts expanding (see Fig.(\ref{fig:E_vs_n})). This shrinkage and expansion indicates the existence of an interaction energy that has exactly same phase boundaries as that of a non-interacting system. Using Eq.(\ref{eq:phaseboundary}), this interaction energy is obtained to be $-2.45$ for $\lambda(x)=(x-0.5)^2+0.25$ (see Fig.(\ref{fig:phaseline}(a)) and the current in system with $E=-2.45$ is higher than its non-interactive counterpart.
Furthermore, the phase diagram in $(n-\lambda_{min})$ plane for interaction energy $E=-2.45$ matches well with $E=0$ when $\lambda(x)=(x-0.5)^2+\lambda_{min}$ (see Fig.(\ref{fig:phaseline}(b)).

\subsection{$E<E_c$}
When interaction energy is less than $E_c$, we obtain three distinct extreme points of current-density relation (Eq.(\ref{eq:Jcmf})): $0.5$, $\rho_{c_1}$ ($<0.5$) and $\rho_{c_2}$ ($>0.5$), where $\rho_{c_1}$ and $\rho_{c_2}$ are the roots of Eq.(\ref{eq:criticalpoints2}). Furthermore, $J$ achieves local maximum at $\rho_{c_1}$ and $\rho_{c_2}$, and local minimum at $0.5$.

To investigate the effect of interactions, we inspect the phase diagram in $n-\lambda_{min}$ plane. Utilizing CMF approach, similar to the case of $E>E_c$, we obtain three different phases, namely, $LD$, $HD$ and $S_{MC}$.
\begin{figure}[h]
\centering
\includegraphics[height=2.5in]{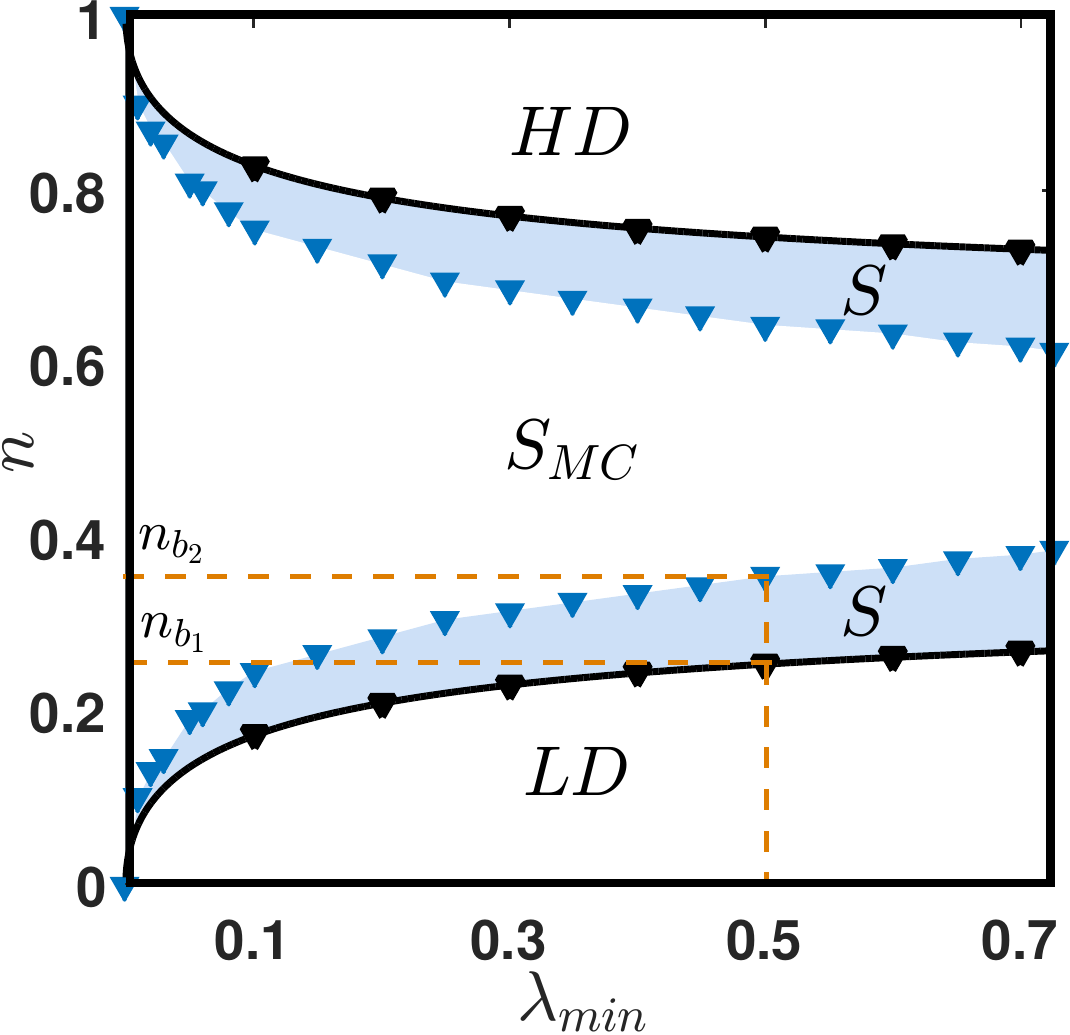}
 \caption{Phase diagram for $\lambda(x)=(x-0.5)^2+\lambda_{min}$ and $E=-4$. Solid line and symbols denote CMF results and MCS results, respectively. CMF predicts three different phases whereas MCS results shows four distinct phases. $n_{b_1}$ and $n_{b_2}$ denote the boundaries between phases for $\lambda_{min}=0.5$. }
 \label{fig:newphase}
 \end{figure}
Densities in these phases, obtained by using Eq.(\ref{eq:rhocmfecrit}), are as follows.
\paragraph*{LD phase:}{The density is smooth throughout the system and is given by $$\rho(x)=\rho_-(x)$$ where $\rho(x)\leq \rho_{c_1}~\forall ~x$ with maxima at $x_0$. At the boundary of $LD$ and $S_{MC}$, $\rho(x_0)=\rho_{c_1}$}.

\paragraph*{$S_{MC}$ phase:}{The density exhibits two shocks, located at $x_0$ and $x_s$ ($>x_0$) for which the profile is given by
\begin{eqnarray}
\rho(x)&=&\begin{cases}  \rho_-(x), &x\leq x_0\\
\rho_+(x),&x_0<x\leq x_s\\
\rho_-(x), &x>x_s.\end{cases}\nonumber
\end{eqnarray}
}
\paragraph*{HD phase:}{The density is smooth throughout the system and is given by $$\rho(x)=\rho_+(x)$$ where $\rho(x)\geq \rho_{c_2}~\forall~x$ with minima at $x_0$. At the boundary of $HD$ and $S_{MC}$, $\rho(x_0)=\rho_{c_2}$}.

\begin{figure}[h]
    \centering
    \includegraphics[height=1.5in]{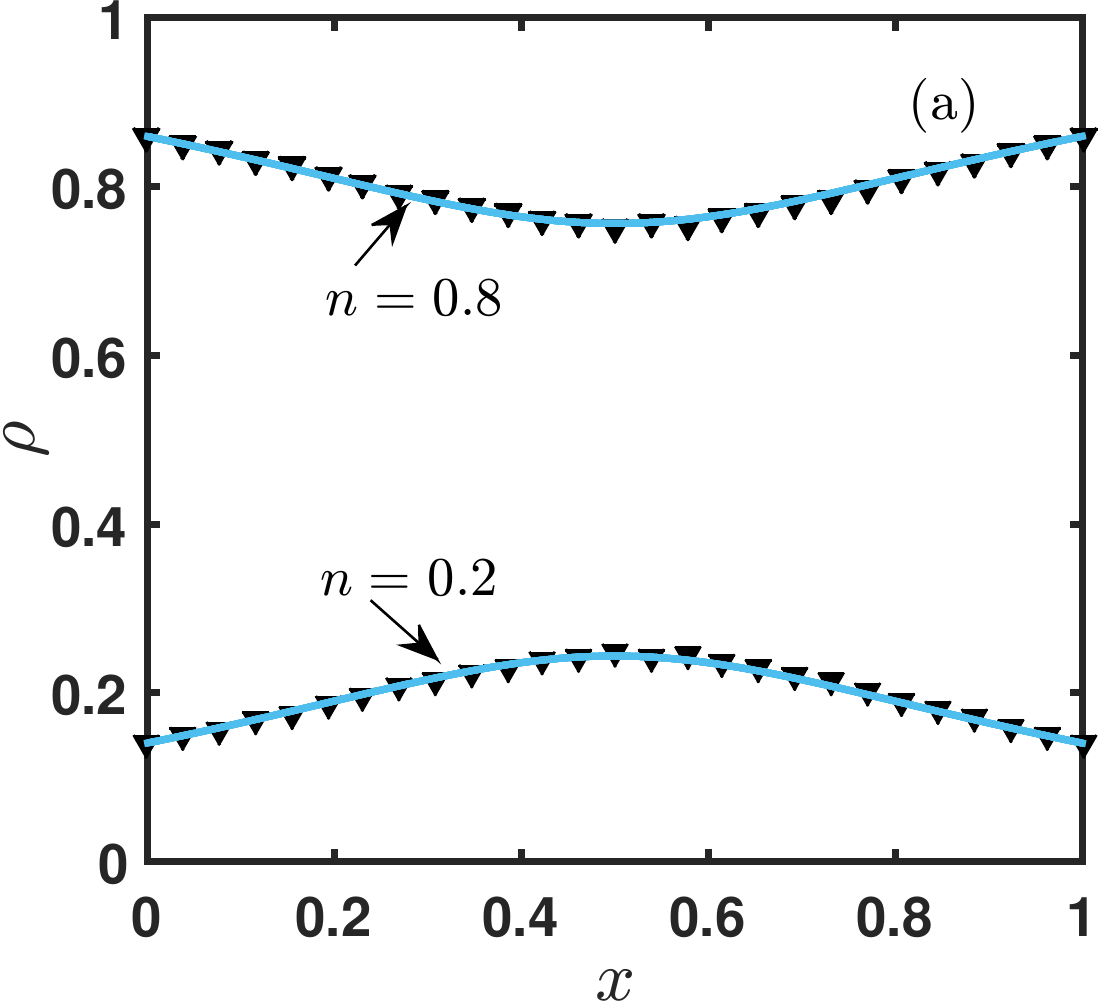}
       \includegraphics[height=1.5in]{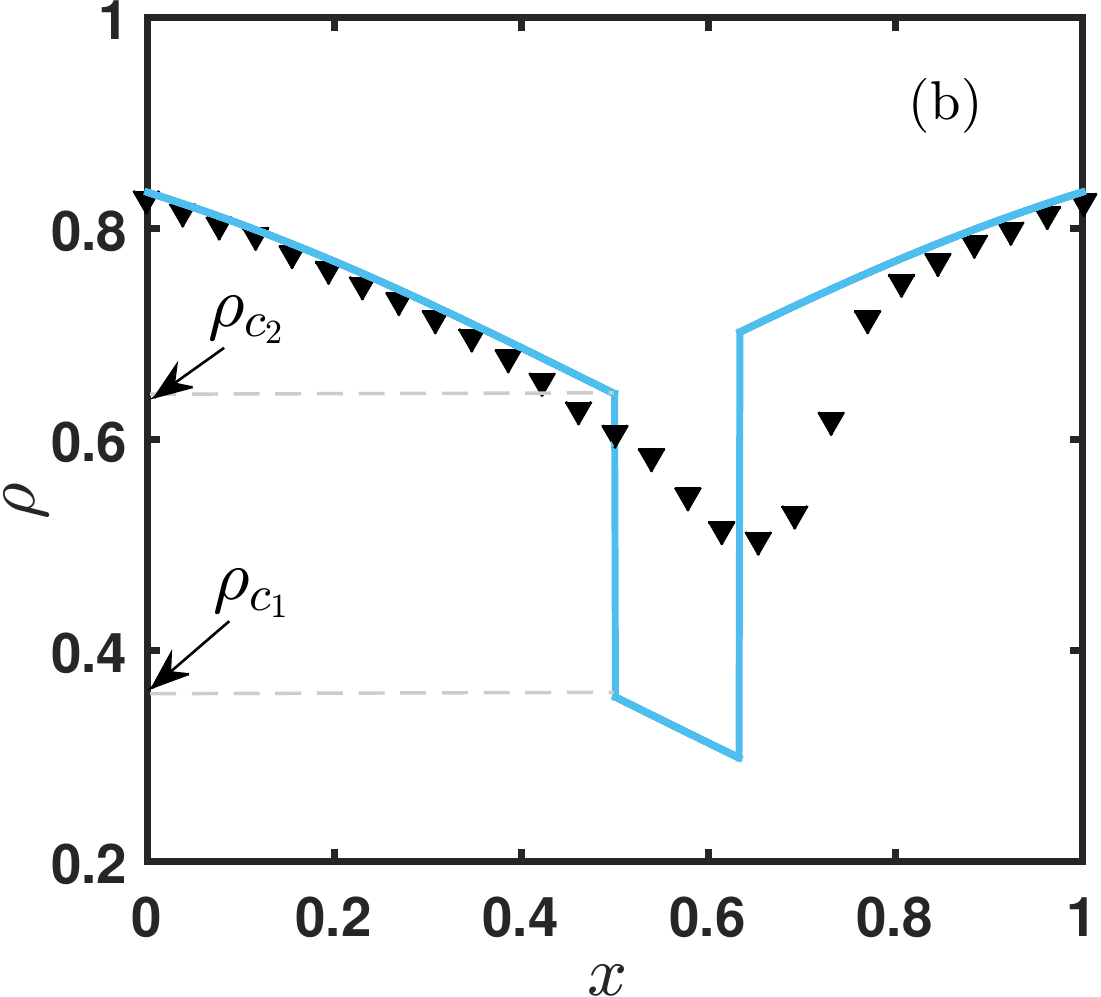}
    \caption{Density profile for $\lambda(x)=(x-0.5)^2+0.5$ and $E=-4$. CMF result is denoted by solid line and MCS result is denoted by symbols. (a) $n=0.2$ and $n=0.8$ (b) $n=0.7$.}
   \label{fig:Ecrit}
\end{figure}

The theoretically obtained density profiles are validated with MCS for specific set of parameters as shown in Fig.(\ref{fig:Ecrit}(a)). Clearly in $HD$ and $LD$ phases, the density profiles are in good agreement. However, in $S_{MC}$ phase, MCS reveals one shock which is in contrast to the theoretical findings which predicted the presence of two shocks (see Fig.(\ref{fig:Ecrit}(b))). Furthurmore, MCS predicts that shock region can be divided into two distinct phases depending upon the position of critical point ($\rho=0.5$). In one phase,  position of critical point is not fixed,whereas in the other phase, the critical point is fixed at $x_0$ i.e., $\rho(x_0)=0.5$ which has characteristics similar to the $S_{MC}$ phase obtained when $E>E_c$. This feature has not been captured by the theoretical finding. We further analyze the shock phase in detail.
\subsubsection*{\textbf{Breakdown of CMF approach and analysis of shock phase}}
As discussed above, CMF and MCS results do not agree largely in shock phase. The discrepancy is probably due to the non-homogeneity in density profile that might have increased the correlations which are not captured by proposed CMF theory. This sort of mismatch is also reported in Ref.\cite{midha2018effect}. Henceforth we use MCS for analyzing the shock phase.

\begin{figure}[h]
    \centering
    \includegraphics[height=2.5in]{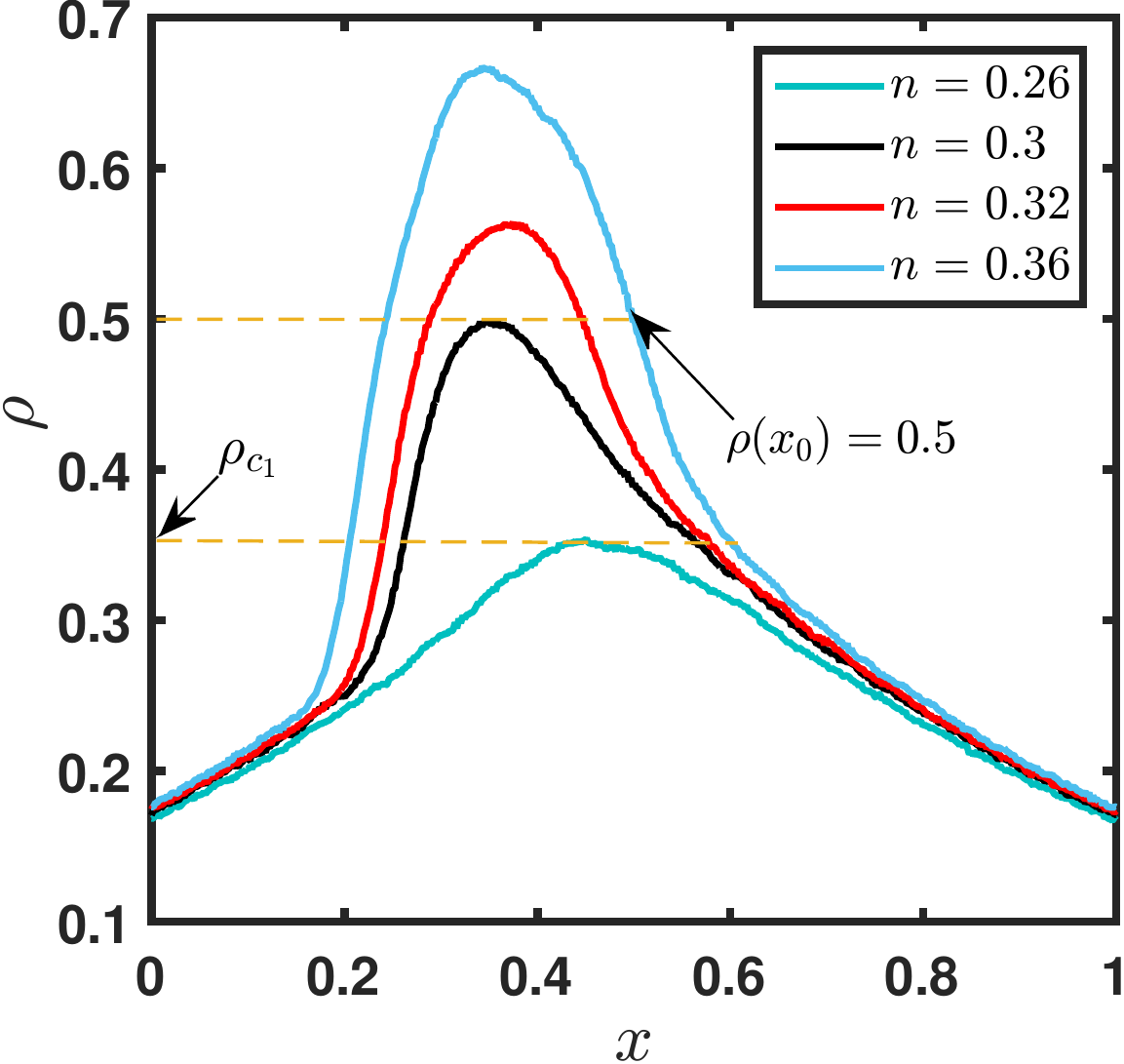}
    \caption{Density profiles for different values of $n$ in $S$ phase where $\lambda(x)=(x-0.5)^2+0.5$ and $E=-4$. Here, shifting of critical value from $x_0$ is clearly visible.}
   \label{fig:SS2}
\end{figure}

Now, we inspect the phase transitions for a fixed $\lambda_{min}$ by varying $n$. To analyze the transition from $LD$ to shock phase, we use the following notations for the specific values of $n$ that separate the distinct phases: $n_{b_1}$ and $n_{b_2}$ as shown in Fig.(\ref{fig:newphase}). As $n$ increases from $n_{b_1}$, we observe that the position of $\rho_{c_1}$ shifts from $x_0$  to $x^*$ $(>x_0)$  and density profile contains only one shock (see Fig.(\ref{fig:SS2})). This is in contrast to findings reported in Ref.\cite{banerjee2020smooth} wherein such a shift is not observed. The shifting continues until $n=n_{b_2}$ and in that stage, $\rho(x_0)$ attains $0.5$. It is also noticed that the current in the system also increases with $n$ while $n_{b_1}< n<n_{b_2}$. This clearly implies that even when the system is in shock phase for $n_{b_1}< n<n_{b_2}$, it does not attain maximal current (see Fig.(\ref{fig:Jvsn})). We denote this phase by $S$. When $n$ is increased further, the position at which the density profile achieves $\rho_{c_1}$ and $0.5$ become fixed and the current in the system also attains a constant value (see Fig.(\ref{fig:Jvsn})). This phase is denoted by $S_{MC}$. Transition from $S_{MC}$ to $HD$ through $S$ can be understood in similar lines.

\begin{figure}[h]
    \centering
    \includegraphics[height=2.5in]{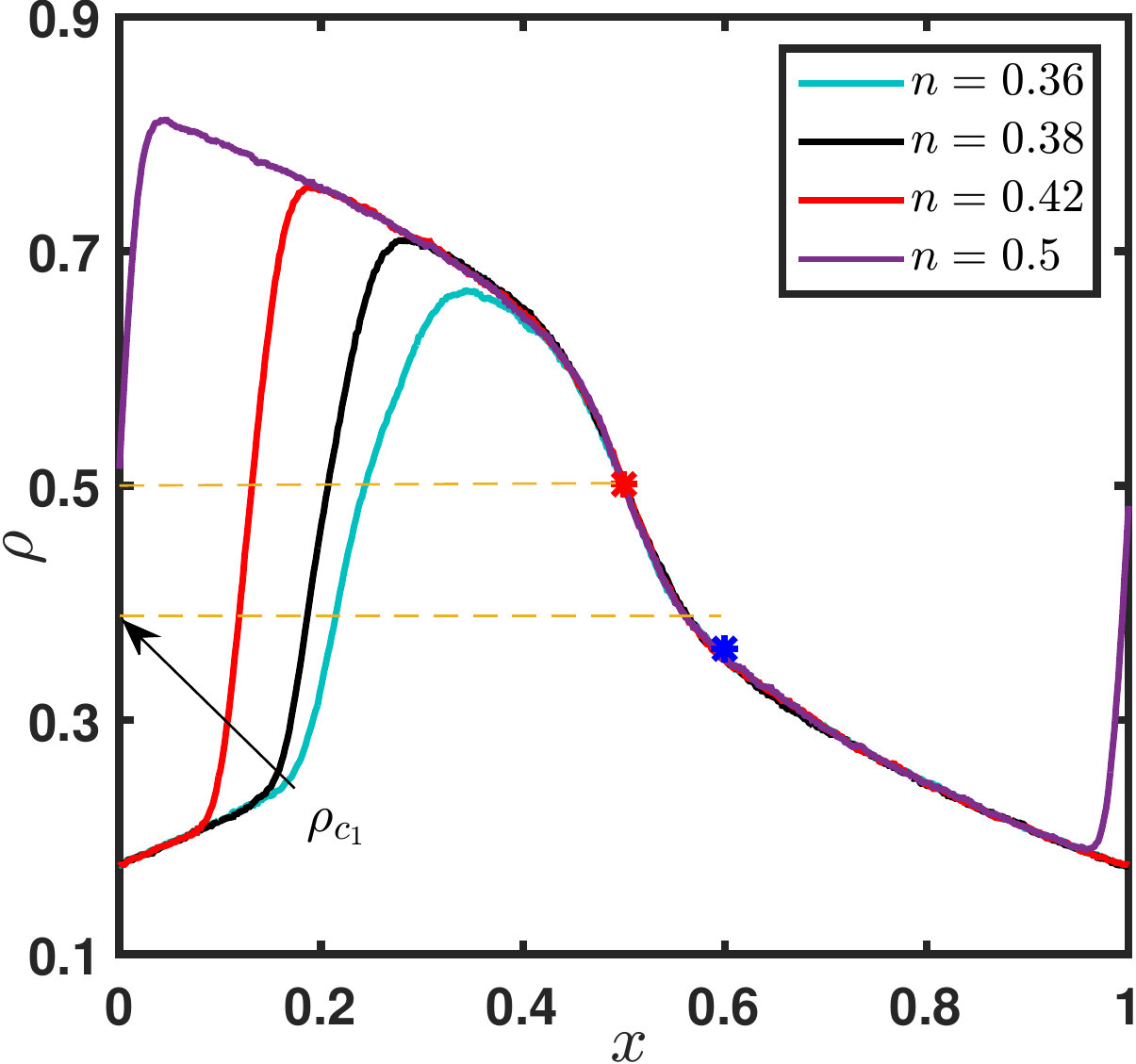}
    \caption{Density profiles for different values of $n$ in $S_{MC}$ phase where $\lambda(x)=(x-0.5)^2+0.5$ and $E=-4$. Asterisk shows that $\rho(x_0)=0.5$, whereas blue asterisk denotes location of $\rho_{c_2}$. Clearly, the location of critical points are fixed.}
   \label{fig:maximal_current}
\end{figure}

Thus, depending on current, we observe two types of phases involving a shock:
\begin{enumerate}[label=(\roman*)]
\item when current varies with number of particles while displaying a shock in density profile.
\item when current remains constant and density exhibits a shock.
\end{enumerate}
Therefore, in the phase diagram in $n-\lambda_{min}$ plane, four distinct phases, namely,  $LD$, $HD$, $S$ and $S_{MC}$ are observed.
The newly observed phase $S$ has bot been reported earlier \cite{banerjee2020smooth}.
Clearly, the current predicted by CMF and MCS depict excellent agreement in LD and HD phases, whereas they do not match in shock phases (see Fig.(\ref{fig:Jvsn})).  CMF predicts maximal current in both $S$ and $S_{MC}$ phases whereas MCS indicates constant current only in $S_{MC}$ phase.

\begin{figure}[h]
\centering
\includegraphics[height=2.5in]{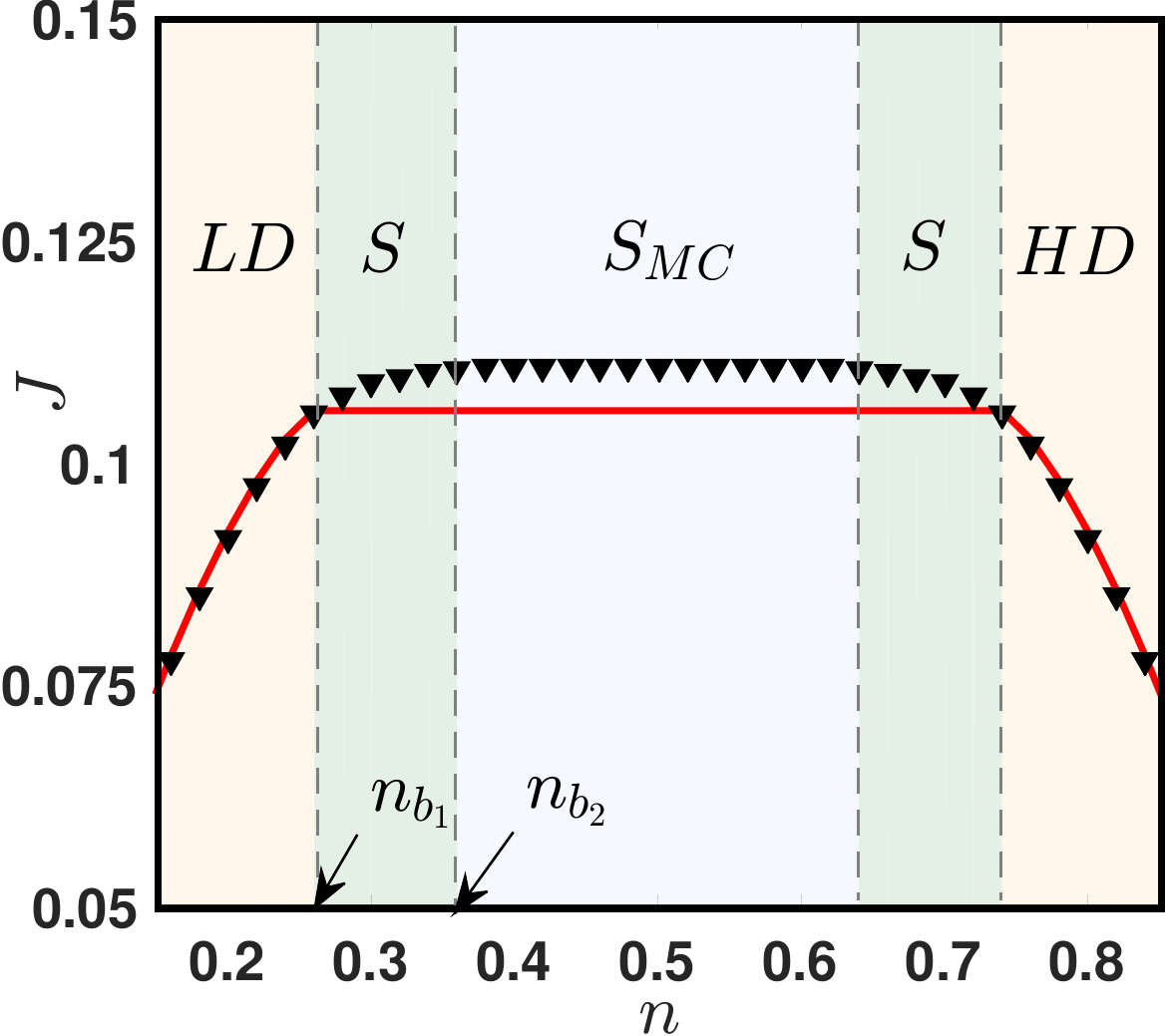}
 \caption{Variation of current $J$ with respect to $n$ for  $\lambda(x)=(0.5-x)^2+0.5$ and $E=-4$. Solid line and symbols denote CMF results and MCS results, respectively.}
 \label{fig:Jvsn}
 \end{figure}


\section{Conclusion}
To summarize, we considered a closed TASEP with interacting particles and quenched hopping rates which are further altered depending upon interaction energy $E$ in a thermodynamically consistent manner. We utilized simple mean field approach which neglects all correlations in the system. It is observed that this theory fails to predict the density profiles and current in the system which is attributed to the correlations that exist in the system. Therefore, to incorporate some correlations, the system is analyzed theoretically using cluster mean field (CMF) approximation. Specifically, we employed two-site CMF and obtained a critical energy $E_c$ such that the current-density relation has one and three extreme points for $E>E_c$ and $E<E_c$, respectively.

It emerges that beyond $E_c$, results in our proposed model have a behavior qualitatively similar to non-interacting system with three distinct phases : $LD$, $HD$ and $S_{MC}$. Nevertheless, it is interesting to note that as interaction energy increases, $S_{MC}$ phase reveals a non-monotonic nature i.e, $S_{MC}$ region decreases followed by an increase after a critical value $E_{c_1}$.
Our observations led us to the fact that there exists an interaction energy for which the boundaries between phases is identical to its non-interactive analogue whereas current is slightly higher in the interactive system.

Below the critical value $E_c$, CMF approach predicts three phases : $LD$, $HD$ and $S_{MC}$. Density profiles obtained by CMF and MCS agree well in $LD$ and $HD$ phase. However, in shock phase, density profile obtained by CMF predicts presence of two shocks, whereas MCS result exhibits only one shock and divides the shock phase into two disjoint phases: $S_{MC}$ and $S$. Thus, MCS reveals four distinct phases: $LD$, $HD$, $S_{MC}$ and $S$. The striking feature that separates new found $S$ from $S_{MC}$ phase is that while the current is not maximal in $S$, it attains its optimal value in $S_{MC}$ phase. The observed distinguished phase, which appears in our system due to interplay between quenched hopping rates and interactions, has not been reported in earlier relevant studies.

Though we have adopted $\lambda(x)=(x-0.5)^2+0.5$ to validate the results with Ref.\cite{banerjee2020smooth}, our methodology is generic and can be used even for discontinuous $\lambda(x)$ (see Appendix \ref{appendix:disc}). The proposed model is not only helpful in explaining the collective dynamics of particles in a conserved environment, but also provided some deep insight into the complex non-equilibrium system.  


\nocite{*}
%

\appendix
\section{}
\label{appendix:mapping}
We obtain here the expressions for density profiles for a non-interactive system comprising of  $l$-mers i.e., particles of size $l$. By setting each hole and each $l$-mer on $L$ sites, as a hole and a monomer, respectively, on another closed lattice with $L\rq{}$ sites, we get a mapping $L\rq{}/L=1-(l-1)\rho_l$, where $L\rq{}=L-(l-1)N$ and $\rho_l$ denotes the density of the system with $N$ $l$-mers and $L$ sites. This leads to the expression of $\rho_l$ as $\rho_l=(L\rq{}/L)\rho_1$, where $\rho_1$ denotes the density of the system with $N$ monomers on $L\rq{}$ sites. Utilizing this mapping gives $J_l=(L\rq{}/L)J_1$. The expressions for steady state current for monomers from \cite{banerjee2020smooth} leads us to the following expression for steady state current
\begin{eqnarray*}
J_l=\lambda(x)\dfrac{\rho(x)\big(1-2\rho(x)\big)}{1-\rho(x)}.
\end{eqnarray*}
\section{}
\label{appendix:disc}
 In our study, the hopping rate function taken into consideration is $\lambda(x)=(0.5-x)^2+\lambda_{min}$ which is a smooth function. It is interesting to note that our analysis is applicable to the system whose hopping rate function might not be differentiable. In fact, our theory also works for a function with discontinuity. However, a shock may appear in the $LD$ and $HD$ phases due to the hopping rate function being discontinuous. We illustrate this in the following subsections.
  \subsection{Finitely Many Discontinuities}
\begin{figure}[h]
 \centering
 \includegraphics[height=1.6in]{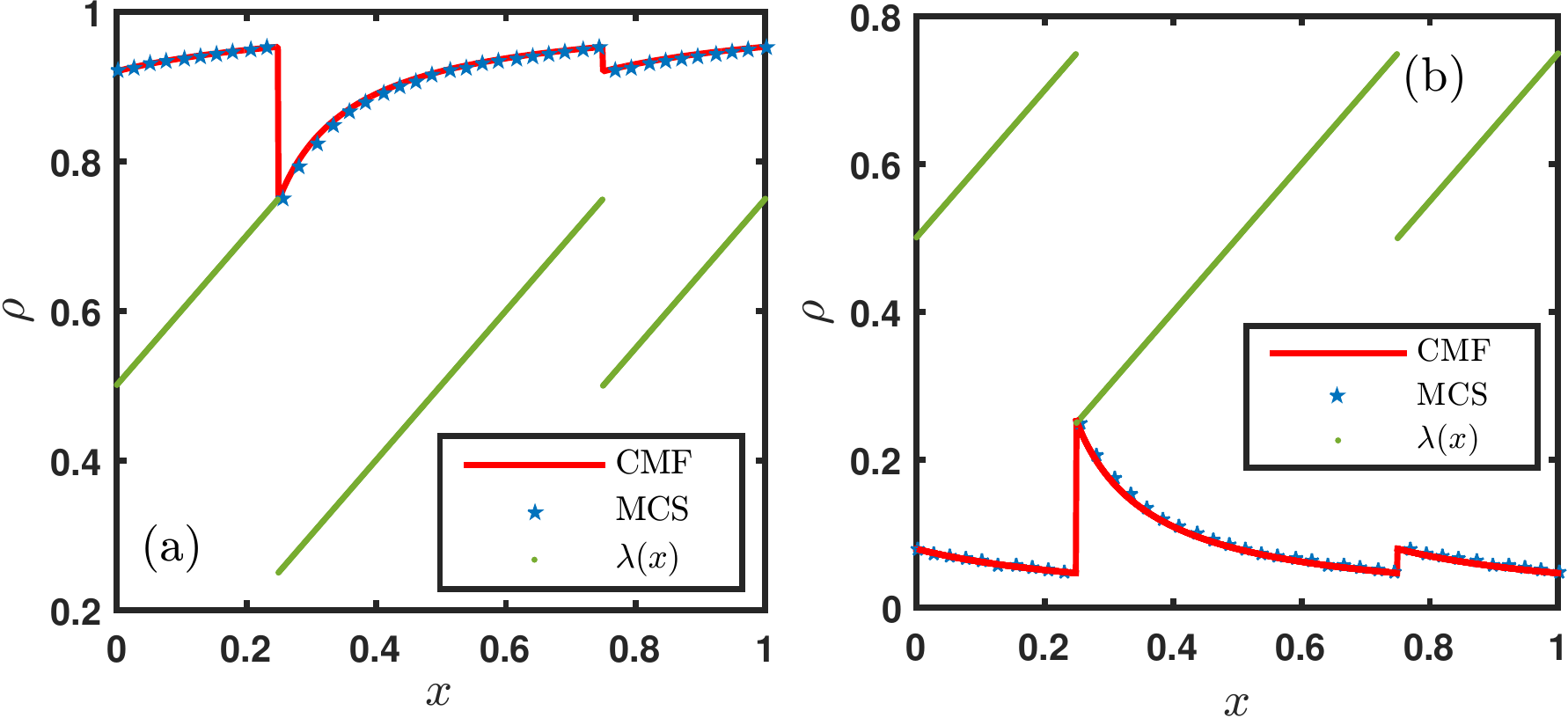}

 \caption{Density profile for $E=1.6$ and (a) $n=0.9$, (b) $n=0.1$. Pentagons and red solid lines denote MCS and CMF results, respectively.}
  \label{fig:twodisc}
 \end{figure}

Fig.(\ref{fig:twodisc}) shows the density profile of a function with two discontinuities. It is evident that system in Fig.(\ref{fig:twodisc}(a)) has HD phase whereas the system in Fig.(\ref{fig:twodisc}(b)) has LD phase. Two shocks are present in the density profile. However, these shocks do not imply shock phase. The density profiles show very good agreement of CMF and MCS.
  \subsection{Infinitely Many Discontinuities}
  \begin{figure}[h]
 \centering
 \includegraphics[height=1.5in]{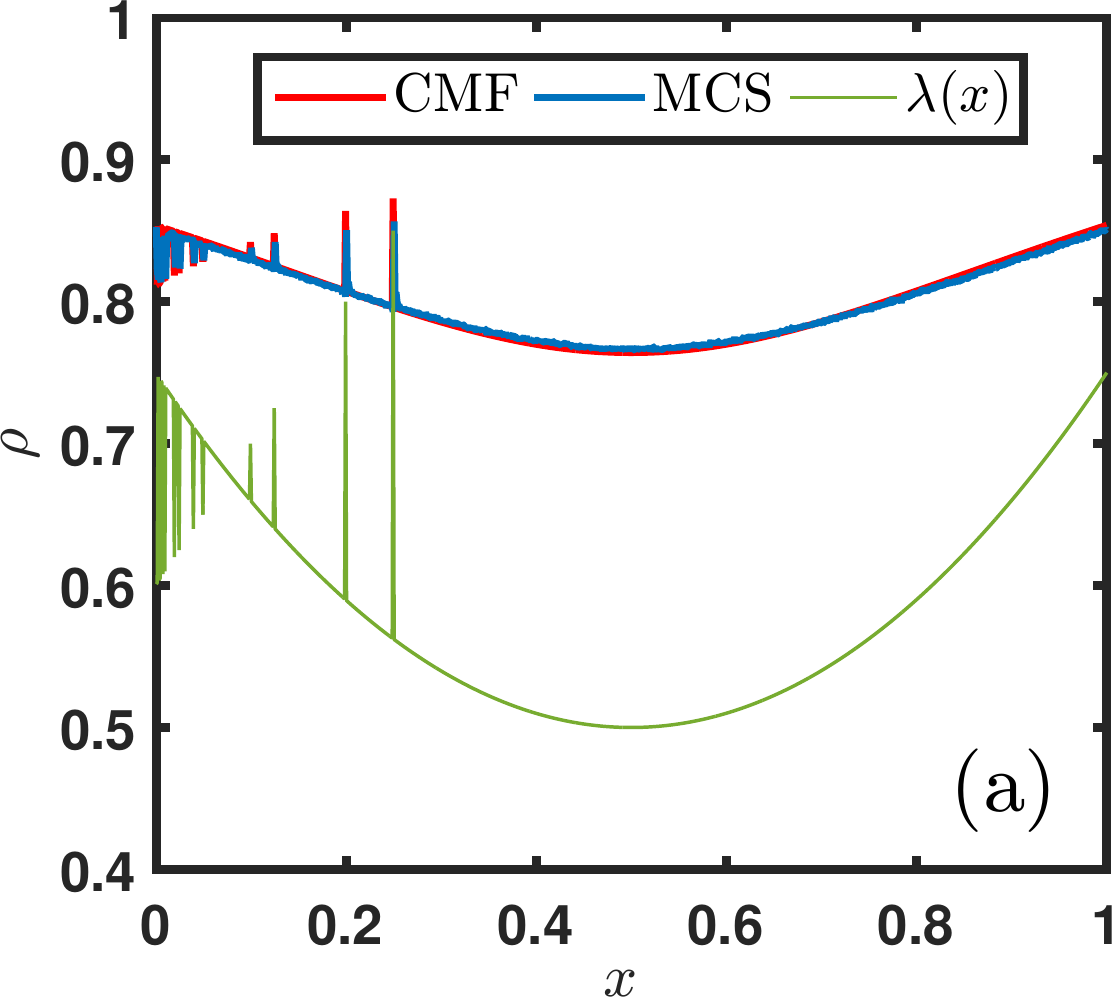}
  \includegraphics[height=1.5in]{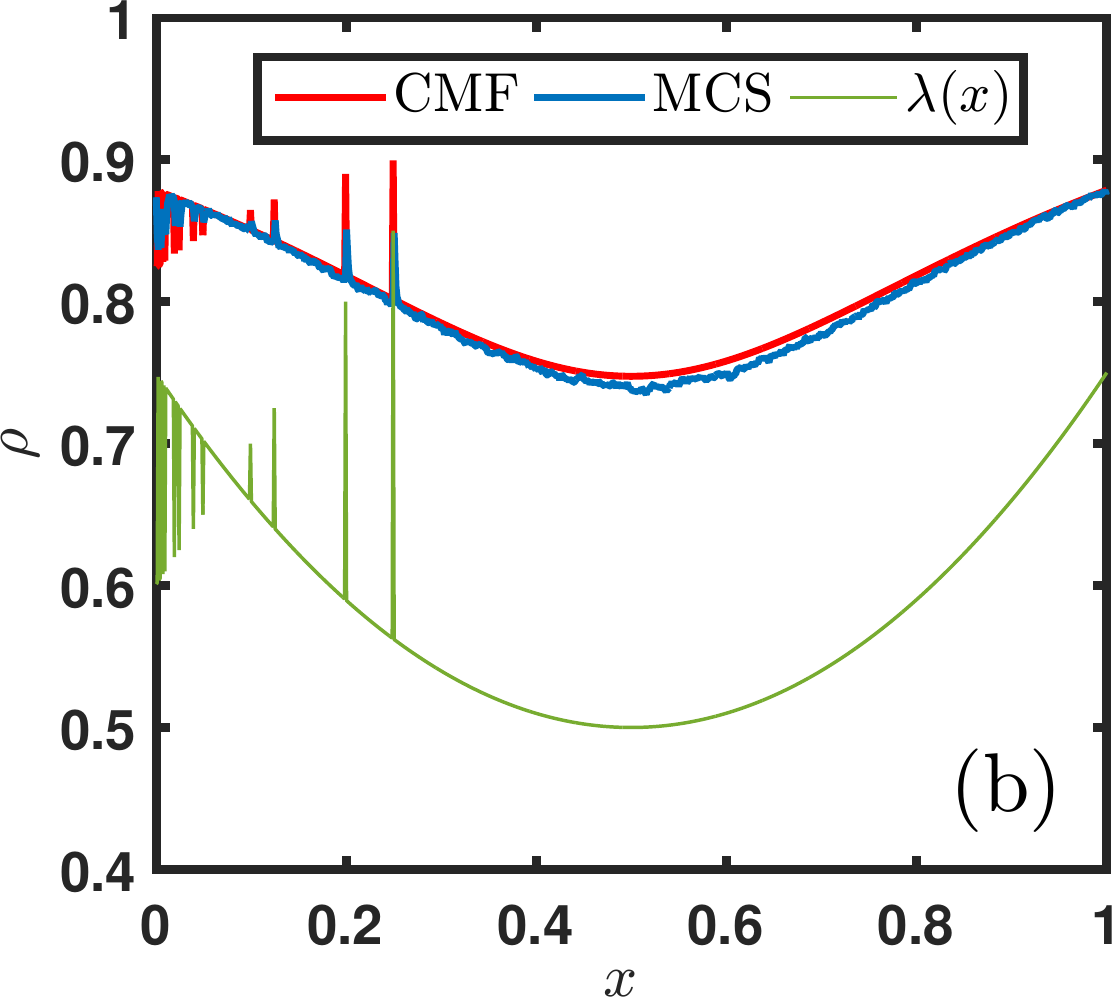}
 \caption{Density profile for (a) $E=1.6$ and $n=0.8$, (b) $E=-1.6$ and $n=0.8$. Blue and red solid lines denote MCS and CMF results, respectively.}
 \label{fig:infdisc}
 \end{figure}
 The question that immediately comes to one\rq{}s mind is: At most how many discontinuities can a function have so that our analysis predict accurate results? As seen in Fig.(\ref{fig:twodisc}), it can be safely said that our system works for finite number of discontinuities. But, can we say something about the behavior if the function has infinitely many discontinuities? To understand such a case, consider a hopping rate function $\lambda(x)$ defined as
 \begin{equation*}
 \lambda(x)\coloneqq
 \begin{cases} x+0.5, &x=\frac{1}{n}, n\in\mathbb{N}\\
 (0.5-x)^2+0.5, &x\neq\frac{1}{n} ,  n\in\mathbb{N}
 \end{cases}
 \end{equation*}

 This function has discontinuities at infinitely many points. We consider a lattice of finite size $L(\gg 1)$. Then, the hopping rates for each site $i$ is calculated as $\lambda_i=\lambda(\frac{i-1}{L})$. However, it is worth noting that these $\lambda_i$ can also be represented as
 \begin{equation*}
 \lambda(x)\coloneqq
 \begin{cases} x+0.5, &x=\frac{1}{n}, n\in \{1,2,\cdots,L\}\\
 (0.5-x)^2+0.5, &x\neq\frac{1}{n}, n\in \{1,2,\cdots,L\}
 \end{cases}
 \end{equation*}
 which again has a finite number of discontinuities and hence, CMF and MCS density profiles match (See Fig(\ref{fig:infdisc})). This is due to the fact that our function with infinite discontinuities reduces to its analogue having finite discontinuities and for such a function, our theory works. In applications, we have finite number of sites which naturally imply that we shall never come across such a $\lambda(x)$.
\end{document}